\documentclass[12pt]{article}
\usepackage{graphicx} 
\usepackage{color}
\usepackage{hyperref}
\usepackage{ifthen}
\usepackage{ulem}
\usepackage{amssymb,natbib,amsthm}
\usepackage{amsmath}
\usepackage{centernot}

\newtheorem{comment}{\color{cyan}Comment}

\theoremstyle{definition}

\theoremstyle{plain}

\usepackage{marginnote}
\usepackage{fontawesome5}

\newcommand{\delText}[1]{{\marginpar{\vspace{-0.15in}\color{red}\Large$\checkmark$}}{\color{red} \sout{#1}}}
\newcommand{\addText}[1]{{\marginpar{\vspace{-0.15in}\color{red}\Large$\checkmark$}}{\color{blue} #1}}

\newcommand{\Weak}{Transformational}
\newcommand{\articulation}{articulation}
\newcommand{\Articulation}{Articulation}
\renewcommand{\articulation}{exploration}
\renewcommand{\Articulation}{Exploration}

\newcommand{\Supplementary}{Supplementary}

\newcommand{\weak}{{exploratory and transformational}}

\usepackage[sectionbib]{bibunits}

\newcommand{\blind}{0}

\title{
Towards Strong AI: Transformational Beliefs \\
and Scientific Creativity
}
\if0\blind
\author{Samuel J. Eschker \;\;\; and \;\;\; Chuanhai Liu \\
Department of Statistics, Purdue University}
\fi

\if1\blind
\author{
\vspace{0.3in}
}
\fi

\newbool{is_arXiv}
\booltrue{is_arXiv} 
\providebool{is_arXiv}

\begin{document}

\def\spacingset#1{\renewcommand{\baselinestretch}%
{#1}\small\normalsize} \spacingset{1} 

\ifbool{is_arXiv}{
\renewcommand{\Supplementary}{Appendix}
\spacingset{1.25} 
\renewcommand{\addText}[1]{#1}
\renewcommand{\delText}[1]{}
}{
\spacingset{1.5} 
}


\maketitle

\begin{quote}
\textit{
\noindent
The zenith of human intelligence is often portrayed as the ability to create, and to create radically new and/or surprising things.
}

    \hfill{--- Geraint A. \cite{wiggins2006preliminary}}
\end{quote}

\begin{abstract}
Strong artificial intelligence (AI) is envisioned to possess general cognitive abilities and scientific creativity comparable to human intelligence, encompassing both knowledge acquisition and problem-solving. While remarkable progress has been made in weak AI, the realization of strong AI remains a topic of intense debate and critical examination. In this paper, we explore pivotal innovations in the history of astronomy and physics, focusing on the discovery of Neptune and the concept of scientific revolutions as perceived by philosophers of science. Building on these insights, we introduce a simple theoretical and statistical framework of weak beliefs, termed the Transformational Belief (TB) framework, designed as a foundation for modeling scientific creativity. Through selected illustrative examples in statistical science, we demonstrate the TB framework's potential as a promising foundation for understanding, analyzing, and even fostering creativity --- paving the way toward the development of strong AI. We conclude with reflections on future research directions and potential advancements.
\end{abstract}

\noindent
{\it Key Words}:
%
Computational creativity;
Chain-of-thought;
Chain-of-verification;
Inferential models;
Generalized logic of science;



\newpage
\ifbool{is_arXiv}{}{
\spacingset{1.9} 
\addtolength{\textheight}{.5in}%
}


\section{Introduction}\label{s:intro}


This paper is the first in a series of three by the authors, exploring statistical reflections on strong AI. Strong AI, as envisioned here, would possess general cognitive abilities and scientific creativity akin to human intelligence, enabling it to approach knowledge acquisition and problem-solving. The other two papers in this series focus on \textit{individual cognition} and \textit{scientific reasoning}, respectively. This explains the title of the present article.
Here, we consider scientific creativity from a statistical perspective. For a concrete discussion at the foundational level, we shall focus on scientific inquiry, where science is simply meant to gain knowledge from experience or experiments \citep[ss, {\it e.g.},][]{newton1718opticks,martin2015inferentialbook}. As it is often the case that science inquiry is dynamic \citep{popper_logic_2005,kuhn1970structure}, a simple but proper statistical setting can be written as 
\begin{equation}\label{eq:usual-setting}
(\Omega_\tau, D_\tau, M_\tau, \Theta_\tau)
\end{equation}
where $\tau$ indices the dynamic state such as sample size and time, $\Omega_\tau$ the world or environment of interest from which the observed data $D_\tau$ were collected, and $M_\tau$ the model with the space $\Theta_\tau$ of unknown parameters. Perhaps, it can be argued that the success of deep learning (DL) is largely due to its flexibility for a $\tau$-dependent or dynamic approach to modeling data structures. A recent relevant discussion, with the focus on parameter estimation, is provided in \cite{jiang2024estimation}. 

Although somewhat special with respect to the complexity of the real world of scientific inquiry, the statistical setting (\ref{eq:usual-setting}) is deemed adequate to interpret the current logic foundations of weak AI. Here, this paper considers the creativity aspect of strong AI. While it has been playing a fundamental role in scientific inquiry from ancient times, creativity has appeared to be such an elusive concept that it is hitherto difficult to have a well-accepted definition. 
For example, in their review paper on computational creativity, 
\cite{carnovalini2020computational} noticed that researchers 
analyzed over 200 of definitions of creativity in literature. 
In the broader context of scientific discovery, philosophers of science have had unsettled debates for centuries on the closely related concept of {\it scientific discovery}, which can be viewed as the processes and products of scientific creativity.  For instance, \cite{sep-scientific-discovery} wrote:
\begin{quote}\it
Philosophical discussions of scientific discovery have been intricate and complex because the term “discovery” has been used in many different ways, both to refer to the outcome and to the procedure of inquiry. In the narrowest sense, the term ``discovery'' refers to the purported ``eureka moment'' of having a new insight.  
\end{quote}
The term of ``eureka moment'' or happy moment in the quote marks its importance in the ultimate definition of creativity consistent with our common sense.

As difficult as it may be, a quantitatively meaningful definition of creativity might not be possible in general. However, we strive for a narrow but precise definition of scientific creativity, particularly in the context of scientific discovery where the existing solutions have been found questionable in terms of either validity or efficiency. Our definition of scientific creativity is formulated into a somewhat simplest possible statistical framework of creativity based on both a study of a selected list of great innovations in the history of science in Section \ref{s:motivating} and our understanding of the perspective of philosophers of science. This framework is formulated within the context of dynamics for scientific discovery and can be summarized briefly as follows. 
Firstly, we expand the special setting (\ref{eq:usual-setting}) into a general setting of scientific discovery at the dynamic state $\tau$: 
\begin{equation}\label{eq:general-setting}
(\Omega_\tau, D_\tau, M_\tau, \Theta_\tau) \stackrel{\mbox{\scriptsize Create}}{\longrightarrow} (\Omega_{\tau'}, D_{\tau'}, M_{\tau'}, \Theta_{\tau'})
\end{equation}
where $\Omega_{\tau'}$ is a new world or population made of the original $\Omega_\tau$ and potentially additional some auxiliary world $\Omega_{\mbox{mis}}$, $D_{\tau'}$ is the new data, and $M_{\tau'}$ is the new model with the space of $\Theta_{\tau'}$ to be considered to address the observed discrepancy in the scientific inquiry. 
Secondly, creative ideas are entertained iteratively or in parallel, according to some iteration among the following three steps:
{\it Creation} constructing $(\Omega_{\tau'}, D_{\tau'}, M_{\tau'}, \Theta_{\tau'})$ in (\ref{eq:general-setting}); {\it \Articulation} entering a necessary stage that \cite{kuhn1970structure} calls \textit{normal research}; and {\it Evaluation} comparing the {\it predicted} against the {\it observed} as a way of evaluation, confirmation, or verification.
When successful, creative solutions are found in the sense that can be considered as the purported ``eureka moment'' of having a new insight. Within this context, our scientific creativity is defined as the transforming procedure of Creation subject to the verification by the Evaluation step. In other words, we consider it creative, as it creates a new world to solve the unsolved problem.  We call the above statistical approach the transformational belief (TB) framework of scientific creativity, with resulting new TBs obtained with respect to the new world indexed by $\tau'$.


\ifthenelse{1=1}{}{
\begin{comment}[Weak Belief: review of creativity and discovery, motivation, and preview of WB]
\color{red} [How to measure or define creativity?]
{\color{red}TODO: Provide a brief review of the topic based on discussions in the scientific literature.} This is consistent with what is to be proposed here in this paper:
\begin{quote}
\begin{description}
    \item[Weak Belief] ``wildly generating lots of crazy possibilities''; and
    \item[Confirmation/assessment] ``(carefully) narrowing in on just the useful ones''.
\end{description}
\end{quote}
{\color{red} Confirmation to be discussed in this paper is to follow the principles in the auto-modeling paper with Yirang Jiang. 
That is, ``usefulness'' is to be measured by comparing the {\it predicted} against the {\it observed}.
}    
\end{comment}
}

The rest of the paper is carried out itself in an inductive fashion and is arranged as follows. Section \ref{s:motivating} develops intuitions for a definition of scientific creativity by studying both a selected list of great innovations in the history of astronomy and physics and the perspectives of philosophers of science on scientific discovery. 
Section \ref{s:WB} presents the TB framework of scientific creativity in a dynamic data-driven environment for problem-solving. 
Section \ref{s:mnms} illustrates TB with a relatively simple artificial example, a many-normal-means problem, showcasing the potential of TB for strong AI, while Section \ref{s:logic-of-science} demonstrates it with the discoveries in the attempts to develop inductive inference, also known as the logic of science \citep[see, {\it e.g.},][]{jaynes_probability_2003}. For the latter, a simple experimental computational evaluation is also considered using ChatGPT \citep{openai2024chatgpt}, a large language model (LLM), with a manually dynamic chain of thought and verification. Section \ref{s:conclusion} concludes with a few remarks for future research.


\section{Historical Discoveries in Natural Science}\label{s:motivating}

In this section, we develop intuitions for our definition of creativity by studying some examples of great discoveries in astronomy and physics. Our exposition aims to provide sufficient details on the discovery of Neptune in Section \ref{ss:neptune} and to summarize our investigations on a list of other great discoveries in the history of astronomy and physics in Section \ref{ss:other-innovations}.  Section \ref{ss:other-innovations} also contains a brief review of the perspectives of philosophers of science on scientific discoveries and revolutions, which more or less agree with our observations.

\subsection{The groundbreaking discovery of Neptune}\label{ss:neptune}

A potential application of \weak~beliefs pertains to a scientist's reasoning process, through which they {\it accept or reject} their theories based on observed evidence. This uses probability theory and has been extensively debated by philosophers for over a century \citep[see, {\it e.g.},][p. 133, \S 5.5]{jaynes_probability_2003}. 
For a specific example, we consider the discovery of Neptune and in a chronological order, our exposition begins with the discovery of Uranus.

\subsubsection{The discovery of Uranus}\label{sss:uranus}
Aside from Earth, five planets — Mercury, Venus, Mars, Jupiter, and Saturn — are easily visible to the naked eye and have been known since ancient times. Uranus was the first planet
discovered with the aid of a telescope by William Herschel on March 13, 1781, while conducting
a systematic sweep of the contents of the night sky. 
Initially, he believed that he had found a comet because the object appeared to move relative to the stars. However, further observations by Herschel and other astronomers revealed that the object had a nearly circular orbit around the Sun at a distance about twice that of Saturn (see the Titius-Bode law \citep{gregory1715elements}),
suggesting that it was a planet rather than a comet, which would have had a highly elliptical orbit. 

Herschel's finding extended the boundaries of the solar system and marked a significant advancement in astronomical research. For example, earlier star catalogs revealed that Uranus had been observed 20 times before its identification as a planet in 1781, dating back as early as 1690, but it was mistakenly identified as a star. Even more intriguingly, in 1821, French astronomer Alexis Bouvard compiled all available observations, spanning a period during which Uranus had traversed about one-third (32/84)
of its orbit, and encountered a significant issue \citep{bouvard1821tablesEnglish}. Despite accounting for the gravitational influences of the giant planets Jupiter and Saturn, he was {\it unable to reconcile the observed data with predictions} based on the Newtonian theory, laws of motion and gravitation. 

\subsubsection{The discovery of Neptune}\label{sss:neptune}

\ifthenelse{1=1}{}{
\begin{figure}[!htb]
\centering
	\includegraphics[width=3.50in]{RJHall-GravitationalPerturbation.pdf}
	\caption{{\color{red} By RJHall - Own work, see Wikipedia. TO DO? Using the actual data before the discovery of Neptune.} At position b, Neptune gravitationally perturbs the orbit of Uranus, pulling it ahead of the predicted location. The reverse is true at a, where the perturbation retards the orbital motion of Uranus.}
\label{fig:perturbation}
\end{figure}
}

The irregularities in the orbit of Uranus led Bouvard to hypothesize some perturbing body. The irregularities, both in the planet's ecliptic longitude and in its radius vector, could have been explained by several hypotheses, including:
\begin{itemize}
    \item[$H1$.] the effect of the Sun's gravity at such a great distance might differ from Newton's description, {\it i.e.}, the Newtonian theory was demolished; 
    \item[$H2$.] the discrepancies might simply be observational error; or
    \item[$H3$.] perhaps Uranus was being pulled, or perturbed, by an  undiscovered planet or multiple planets 
    that are farther away from Uranus \citep{grosser1964search}.
\end{itemize}
Like Bouvard, French astronomer Urbain Jean Joseph Leverrier (1811-1877) and English scholar John Couch Adams (1819-1892) from St John's College, Cambridge, likely regarded $H3$ as more plausible than $H1$, $H2$, and potentially many other hypotheses. This is evident from the immense computational efforts they invested in further pursuing their mathematical search of the hypothetical disturbing planet.
Working within the framework of H3, Adams and Verrier independently reached a predicted position for the hypothesized perturber on the celestial sphere, using essentially the same perturbation theory techniques and the Titius-Bode law. On 24 September 1846, Johann Gottfried Galle and Heinrich Louis d'Arrest at the Berlin Observatory spotted a new planet that is very close to the predicted position by Verrier, who named this planet of the Solar system {\it Neptune}. This is a truly remarkable achievement for the epoch, as
 \cite{airy1846account} wrote (p.121):
\begin{quote}\it
    In the whole history of astronomy, I had almost said in the whole history of science, there is nothing comparable to this. The history of the discoveries of new planets in the latter part of the last century, and in the present century, offers nothing analogous to it.
\end{quote}
\ifthenelse{1=1}{}{
Regarding the second memior on the theory of {\it Urnnus} given by M. Lv Verrier (Jun 1, 1846, {\it Compte Rendu de l'Acad\'{e}mie}),
\cite{airy1846account} wrote (p.131):
\begin{quote}\it
The first part contains the results of a new reduction of nearly all the existing observations of {\bf Uranus}, and their treatment with reference to the theory of perturbations, as amended in the former memoir. After concluding from this reduction that observations are absolutely irreconcilable with the theory, M. Lv. Verrier considers in the second part all the possible explanations of the discordance, and concludes that none is admissible, except that of a disturbing planet exterior to {\bf Uranus}. He then proceeds to investigate the elements of the orbit of such a planet, assuming that its mean distance is double that of {\bf Uranus}, and that its orbit is in the place of the ecliptic.
The value of the mean distance, it is to be remarked, is not fixed entirely by Bode's law, although suggested by it; several considerations are stated which compel us to take a mean distance, not {\bf very} greatly differing from that suggested by the law, but which nevertheless, without the suggestions of that law, would leave the mean distance in a most troublesome uncertainty. The peculiarity of the form which the investigation takes is then explained.
\end{quote}
}

In modern terms, the problem tackled by Adams and Verrier is an inverse problem. In a recent revisit to the problem, \cite{rodriguez2024discovery} recomputed the perturbations induced in the orbit of Uranus by Neptune, using the data from Solar System Dynamics at \href{https://ssd.jpl.nasa.gov}{\tt https://ssd.jpl.nasa.gov}. 
\ifthenelse{1=1}{}{
Concluding their investigation, the authors of \cite{rodriguez2024discovery} wrote:
\begin{quote}\it 
By using a modern, numerical approach of the gravitational N-body problem by means of the Wisdom-Holman map, we were able to successfully study the problem of the irregularities in the orbit of Uranus due to Neptune, revisiting the milestone made by Adams and Le Verrier. The effect of each giant planet on the orbit was also separately analyzed; as was to be expected, the shifts in the position of Uranus due to Jupiter and Saturn were much greater than that caused by Neptune, exceeding even 1◦ in the geocentric equatorial coordinates after a relatively short period of time. Finally, we considered the perturbation that the Pluto-Charon system inflicts on Neptune, sustaining that the reported discrepances in the orbits of the ice giants at the beginning of the last century could not be produced by it. 

\end{quote}
}





\subsubsection{Statistical evidence for patterns of scientific discovery}\label{sss:neptune-stat-summary}
It is remarkably worth noting that the process in the discovery of Uranus begins with an abnormal phenomenon that contradicts the prediction principle: {\it the observed is inconsistent with the expected or predicted}, providing clear evidence that demands further investigation. The same phenomenon occurred again with the discovery of Neptune.
Perhaps more importantly, our observations suggest the possibility of statistical modeling of scientific discovery.

The above observations no doubt shed light on a meaningful definition of scientific creativity. Typically, creative innovations come next when investigators conduct a new scientific investigation to resolve the discovered anomaly. In the case of the discovery of Uranus, astronomers weakened their previous beliefs, and thus established new beliefs by remodeling the observed data. In the case of the discovery of Neptune, the beliefs were transformed, followed by remodeling the orbit data Uranus with missing values, the hypothetical planets. 

\subsection{More examples of great innovations in astronomy and physics}\label{ss:other-innovations}
\ifthenelse{1=1}{}{
\begin{figure}[!htb]
\centering
	\includegraphics[width=4.50in]{Ptolemaic-system.png}
	\caption{{\color{magenta}(Geocentric Model @ Ancient Times)} Although the basic tenets of Greek geocentrism were established by the time of Aristotle (384–322 BC), the details of his system did not become standard. Claudius Ptolemy (c. 100 - c. 170 AD), an Alexandrian mathematician, astronomer, astrologer, geographer, and music theorist, thought the solar system looked like this and standardized geocentrism. See Wikipedia/Geocentric Model.}
\label{fig:ptolemaic-system}
\end{figure}
}

\ifthenelse{1=1}{

Far more historical examples are available than we have had space to exploit here.
The observations of our studies of great discoveries in celestial mechanics and physics from a statistical perspective of their innovations are briefly summarized in \Supplementary~A. Great discoveries all typically start with experiments and observations, build mathematical theories or statistical models, verify the new theories with experiments and new observations, and iterate such a process towards further verification, improvement, and discoveries.

Extensive existing research on the science of creativity and discovery, primarily conducted by philosophers of science, has also been undertaken \citep[see, {\it e.g.},][and references therein]{kuhn1970structure,Aleinikov2013,sep-scientific-discovery}. Their perceived general structure or pattern of such activities, particularly in natural science, exhibits similarities to our observations in this section. From a statistical perspective, we formulate in the next section a framework of scientific discovery in environments characterized by such a common process.

}{
Far more historical examples are available than we have had space to exploit here.
In this section, we briefly summarize the observations of our studies of great discoveries in celestial mechanics and physics from a statistical perspective of their innovations. The first is the invention of the {\it geocentric model} that hypothesizes that the Sun, Moon, stars, and planets all orbit Earth. 
It was the predominant description of the cosmos in many ancient European civilizations, such as those of Aristotle (384–322 BC) in classical Greece and Ptolemy in Roman Egypt, as well as during the Islamic Golden Age. Claudius Ptolemy (c. 100 - c. 170 AD), an Alexandrian mathematician, astronomer, astrologer, geographer, and music theorist, thought the solar system looked like this and standardized geocentrism. From a statistical perspective of the underlying creativity, we note that this model was built on two observations \citep[][pp. 5-20  and Wikipedia, 10/23/2024]{kuhn1992copernican}: 
\begin{quote}\it First, from anywhere on Earth, the Sun appears to revolve around Earth once per day. While the Moon and the planets have their own motions, they also appear to revolve around Earth about once a day. The stars appeared to be fixed on a celestial sphere rotating once each day about an axis through the geographic poles of Earth. Second, the Earth seems to be unmoving from the perspective of an earthbound observer; it feels solid, stable, and stationary.

\end{quote}

The geocentric model dominated for centuries until the discovery of the heliocentric system, known as the {\it Heliocentric Hypothesis}, of the Greek astronomer and mathematician Aristarchus of Samos (c.310 – c. 230 BC). The heliocentric model places all of the then-known planets in their correct order around the Sun. 
With the invention of the telescope in 1609, {\it observations} made by Galileo Galilei led him to the thought that while this observation was incompatible with the Ptolemaic system, it was a natural consequence of the heliocentric system. Continued {\it observations} and analysis led to variations of
heliocentric system, where planets orbit in perfect circles.

Johannes Kepler (1571 – 1630) 
analyzed Tycho Brahe's 
famously accurate {\it observations} and afterwards constructed his three laws in 1609 and 1619: 
the orbit of a planet is an ellipse with the Sun at one of the two foci;
a line segment joining a planet and the Sun sweeps out equal areas during equal intervals of time; and
the square of a planet's orbital period is proportional to the cube of the length of the semi-major axis of its orbit.     
The application of these laws dramatically improved the accuracy of celestial {\it observations} and {\it predictions}. Violations of Kepler's laws, indicated by inconsistency between the observed and predicted for Mercury, 
have been used to explore more sophisticated models of {\it gravity}, such as general relativity. While Newton's laws generalize Kepler's laws, most problems related to the periods of orbits are still best solved using Kepler's laws, since they are simpler.

Great discoveries were also found in classical physics. For example, the 17th-19th centuries saw Newton's laws of motion and universal gravitation (1687), 
electromagnetism, and thermodynamics.
In April 1820, Hans Christian Ørsted observed that an electrical current in a wire caused a nearby compass needle to move. 
This discovery represented a major step toward a unified concept of energy. This unification, which was observed by Michael Faraday, extended by James Clerk Maxwell, and partially reformulated by Oliver Heaviside and Heinrich Hertz, is one of the key accomplishments of 19th-century mathematical physics \citep{darrigol2003electrodynamics}. 
Scientists like Sadi Carnot, Lord Kelvin, and Rudolf Clausius laid the foundation for the study of heat, energy, and entropy, which had profound implications for understanding physical processes. Discoveries can often be well understood by tracing their history. 
Tracing their history, one would see that like the Neptune discovery, geocentric system, heliocentric system, Kepler's laws, Newton's laws, and Einstein's theoy of special and general relativity, such great discoveries all start with experiments and observations, build mathematical theories or statistical models, verify the new theories with experiments and new observations, and iterate such a process to further verification, improvement, and discoveries. 

{

}

\ifthenelse{1=1}{
Extensive existing research on the science of creativity and discovery, primarily conducted by philosophers of science, has also been undertaken \citep[see, {\it e.g.},][and references therein]{kuhn1970structure,Aleinikov2013,sep-scientific-discovery}. Their perceived general structure or pattern of such activities, particularly in natural science, exhibits similarities to our observations in this section. From a statistical perspective, we formulate a framework of scientific discovery in environments characterized by such a common process in the next section.
}{ 
\subsection{A brief historical review of research on science of creativity}\label{ss:philo-review}
On the {\it science of creativity}, the study of the complex phenomenon of creativity and known as creatology,
\cite{Aleinikov2013} 
wrote:
\begin{quote}\it
As with any new field of research, the creativity research at a certain moment of time develops into a science. Creating (structuring, designing) a new science is an act of creativity. If scientists create new methods of research, new models, new hypotheses, new theories, new devices, new experiments, and these are all creative acts, then the creation of a science can be considered one of the largest creative acts in the field of science. In the history of science, the founders of new sciences are often referred to as ``fathers,'' such as Gregor Mendel, the father of genetics, or Norbert Wiener, the father of cybernetics.

...

It is generally accepted that research on creativity started with G. Wallas' work in which he dissected the act of creativity into four stages: preparation, incubation, illumination, and verification \cite{wallas1926art}. 

\end{quote}

See \cite{sep-scientific-discovery}'s discussion of \cite{kuhn1970structure}.

}
}


\section{\Weak~Beliefs: a General Framework}\label{s:WB}

\subsection{The prediction principle}

Although the focus here is on scientific creativity, our discussion cannot be independent of the scope of scientific discovery for problem-solving in scientific inquiry. It can arguably be said that one of the first principles of science is the principle that the observed data and the predicted data must be consistent with each other. The general idea can be traced to Isaac Newton (1704, 
\cite{newton1718opticks} and \cite{sep-scientific-discovery}), as seen in his method of analysis:
\begin{quote} \it 
``As in Mathematicks, so in Natural Philosophy, the Investigation of difficult Things by the Method of Analysis, ought ever to precede the Method of Composition. This Analysis consists in making Experiments and Observations, and in drawing general Conclusions from them by Induction, and admitting of no Objections against the Conclusions, but such as are taken from Experiments, or other certain Truths … By this way of Analysis we may proceed from Compounds to Ingredients, and from Motions to the Forces producing them; and in general, from Effects to their Causes, and from particular Causes to more general ones, till the Argument end in the most general. This is the Method of Analysis'' 
\end{quote}
We refer to the fundamental underlying principle as the {\it prediction principle }. This principle was made clear in William Whewell's view, as \cite{snyder2023william} summarized:
\begin{quote} \it 
On Whewell's view, once a theory is invented by discoverers' induction, it must pass a variety of tests before it can be considered confirmed as an empirical truth. These tests are prediction, consilience, and coherence \citep[see][{\it p}. 83-96]{whewell1858novum}. 
These are characterized by Whewell as, first, that ``our hypotheses ought to \textit{fortel} [sic] phenomena which have not yet been observed'' \citep[][{\it p}. 86]{whewell1858novum}; 
second, that they should ``explain and determine cases of a \textit{kind different} from those which were contemplated in the formation'' of those hypotheses \citep[][{\it p}. 88]{whewell1858novum}; 
third that hypotheses must ``become more coherent'' over time \citep[][{\it p}. 91]{whewell1858novum}. 
\end{quote}
Notably, using modern inductive inference in terms of a sound logic of science (see Section \ref{s:logic-of-science}), we can arguably perform Whewell's tests of significance against {\it consilience} and {\it coherence} based on the more fundamental concept of {\it prediction}. 

With the prediction principle, here in this section we consider a narrow but concrete definition of scientific creativity that is summarized the three-step TB framework introduced in Section \ref{s:intro}. 

\subsection{A philosophical perspective}

The three-step idea here is similar to or, to some extent, can even be viewed as a modern logic of science-based renovation of that of scientific discovery of three steps of \cite{whewell1840philosophy}; see the discussion in \cite{sep-scientific-discovery}: the happy thought or `eureka moment', the articulation and development of that thought, and the testing and evaluation of it. While there are diverse opinions of philosophers of science on possible logic of discovery, we focus on a pragmatic approach by considering the context of problem-solving as the setting. 
Philosophically, this is dfferent from W. Whewell, as, for example,  \cite{sep-scientific-discovery} wrote:
   `{\it According to Whewell, the initial step in every discovery is what he called ``come happy thought, of which we cannot trace the origin; some fortunate cast of intellect, rising above all rules. No maxims can be given which inevitably lead to discovery'' (Whewell 1996 [1840]:186). }'
To the extent of simulating happy thoughts, especially in the current prevailing era of generative-AI, this seems to be consistent with the common sense understanding of creativity, {\it i.e.}, generating a vast array of unconventional hypotheses. 
This is also consistent with the idea in the literature that the special logic of discovery is the logic of abductive or ``retroductive'' inferences \citep{hanson1958patterns,hanson2014there,sep-scientific-discovery}, which we refer to in Section \ref{ss:TB-framework} as pragmatic reasoning of ``reverse-engineering''. 

However, it is critical to have deep reflections when it comes to thinking about strong AI. In reality, it is important to observe that having the happy thought in discovery is unlikely to be a wild and capricious guess \citep[see][and references therein for similar opinions]{sep-scientific-discovery}. As in the studies shown in Section \ref{s:motivating}, scientific discoveries often involve noticeable inconsistencies between experience and theory. Actions are then necessary to change beliefs that have been obtained from a 
logic of science based on the current propositions, the data, and the resulting model. This explains the rationale for our proposed approach, called the framework of {\it \weak~beliefs} or, simply, the \textit{transformational belief} (TB) framework. 

\ifthenelse{1=1}{}{
\begin{quote}\cite{sep-scientific-discovery}: \it
    7. Anomalies and the structure of discovery

Many philosophers maintain that discovery is a legitimate topic for philosophy of science while abandoning the notion that there is a logic of discovery. One very influential approach is Thomas Kuhn’s analysis of the emergence of novel facts and theories (Kuhn 1970 [1962]: chapter 6). Kuhn identifies a general pattern of discovery as part of his account of scientific change. A discovery is not a simple act, but an extended, complex process, which culminates in paradigm changes. Paradigms are the symbolic generalizations, metaphysical commitments, values, and exemplars that are shared by a community of scientists and that guide the research of that community. Paradigm-based, normal science does not aim at novelty but instead at the development, extension, and articulation of accepted paradigms. A discovery begins with an anomaly, that is, with the recognition that the expectations induced by an established paradigm are being violated. The process of discovery involves several aspects: observations of an anomalous phenomenon, attempts to conceptualize it, and changes in the paradigm so that the anomaly can be accommodated.

It is the mark of success of normal science that it does not make transformative discoveries, and yet such discoveries come about as a consequence of normal, paradigm-guided science. The more detailed and the better developed a paradigm, the more precise are its predictions. The more precisely the researchers know what to expect, the better they are able to recognize anomalous results and violations of expectations:

    novelty ordinarily emerges only for the man who, knowing with precision what he should expect, is able to recognize that something has gone wrong. Anomaly appears only against the background provided by the paradigm. (Kuhn 1970 [1962]: 65)

Drawing on several historical examples, Kuhn argues that it is usually impossible to identify the very moment when something was discovered or even the individual who made the discovery. Kuhn illustrates these points with the discovery of oxygen (see Kuhn 1970 [1962]: 53–56). Oxygen had not been discovered before 1774 and had been discovered by 1777. Even before 1774, Lavoisier had noticed that something was wrong with phlogiston theory, but he was unable to move forward. Two other investigators, C. W. Scheele and Joseph Priestley, independently identified a gas obtained from heating solid substances. But Scheele’s work remained unpublished until after 1777, and Priestley did not identify his substance as a new sort of gas. In 1777, Lavoisier presented the oxygen theory of combustion, which gave rise to fundamental reconceptualization of chemistry. But according to this theory as Lavoisier first presented it, oxygen was not a chemical element. It was an atomic “principle of acidity” and oxygen gas was a combination of that principle with caloric. According to Kuhn, all of these developments are part of the discovery of oxygen, but none of them can be singled out as “the” act of discovery.

In pre-paradigmatic periods or in times of paradigm crisis, theory-induced discoveries may happen. In these periods, scientists speculate and develop tentative theories, which may lead to novel expectations and experiments and observations to test whether these expectations can be confirmed. Even though no precise predictions can be made, phenomena that are thus uncovered are often not quite what had been expected. In these situations, the simultaneous exploration of the new phenomena and articulation of the tentative hypotheses together bring about discovery.

In cases like the discovery of oxygen, by contrast, which took place while a paradigm was already in place, the unexpected becomes apparent only slowly, with difficulty, and against some resistance. Only gradually do the anomalies become visible as such. It takes time for the investigators to recognize “both that something is and what it is” (Kuhn 1970 [1962]: 55). Eventually, a new paradigm becomes established and the anomalous phenomena become the expected phenomena.

Recent studies in cognitive neuroscience of brain activity during periods of conceptual change support Kuhn’s view that conceptual change is hard to achieve. These studies examine the neural processes that are involved in the recognition of anomalies and compare them with the brain activity involved in the processing of information that is consistent with preferred theories. The studies suggest that the two types of data are processed differently (Dunbar et al. 2007).
\end{quote}
}

\subsection{Abduction, reverse-engineering, and the TB framework}\label{ss:TB-framework}

Our discussion of scientific creativity for strong AI is narrow and thus could be said to be an inquiry of statistical discovery.
Recall that in general, problem-solving is an iterative process until certain conditions implementing the prediction principle are satisfied. Thus, in the context of statistical modeling of scientific creativity, the basic idea can be summarized as: at a simplest level, all your need is to consider {\it reverse-engineering, re-sampling, and re-modeling;}
following Albert Einstein:
\begin{quote}
\textit{It can scarcely be denied that the supreme goal of all theory is to make the irreducible basic elements as simple and as few as possible without having to surrender the adequate representation of a single datum of experience.}
\end{quote}
This is often quoted as ‘Everything should be made as simple as possible, but not simpler',
`On the Method of Theoretical Physics', lecture delivered at Oxford, 10 June 1933; See \cite{ratcliffe2014oxford}.

Technically, our jump point is therefore the setting for such a dynamic environment introduced as (\ref{eq:usual-setting}), where
for each iteration, we have an unsolved problem raised by verification of the observed and predicted. With this setting, we discuss the three steps of each iteration in detail below.

\begin{description}
\item{\large \it The creation step.}\label{ss:creation}
Given the unsolved problem, creative approaches relay on the appropriate experience or data to be created to solve the problem. This can include re-sampling and retrospective reconstruction (or reverse-engineering) of data available for the previous analysis. It may also include approaches of retrospective reconstruction or reverse-engineering to select or remove observations. The goal is that the selected or remaining data with the new data, denoted by $D_{\tau'}$, would be used via adequate remodeling to produce valid inference. 
Consequently, we have the population underlying $D_{\tau'}$, which is the resulting population or the explicitly targeted population that represents
the environmental context of the problem to be solved. This leads to
a dynamical modification of (\ref{eq:usual-setting}) to (\ref{eq:general-setting}) and, thereby, the \textit{transformed beliefs}.
\ifthenelse{1=1}{}{, restated as follows:
\begin{equation}\label{eq:general-setting-02}
(\Omega_\tau, D_\tau, M_\tau, \Theta_\tau) \stackrel{\mbox{\scriptsize Create}}{\longrightarrow} (\Omega_\tau', D_\tau', M_\tau', \Theta_\tau')
\end{equation}
where $\Omega_\tau'$ is a new world or population made of the original $\Omega_\tau$ and some auxiliary world $\Omega_{\mbox{mis}}$, $D_\tau'$ is the new data, and $M_\tau'$ with the space of $\Theta_\tau'$ to be considered to address the observed discrepancy in the scientific inquiry. }

\item{\large \it The \articulation~step.}\label{ss:production}
This step, also known as the articulation and development step, aims to advance research based on the new theory or model by formulating and disseminating mathematical or statistical hypotheses about quantities of interest, grounded in scientific logic. It facilitates ongoing research, leading to refined beliefs, including more broadly exploratory ones. When necessary, these probabilistic findings can lead to the subsequent evaluation step with numerical inputs.



\item{\large \it The evaluation step.}\label{ss:evaluation}
    This step, also known as the verification step, applies the prediction principle to compare {\it predicted} against {\it observed} as a way of evaluation, confirmation, or verification. At a technical level, this step involves formulating assertions or hypotheses on the validity of the current model or system of beliefs and generating uncertainty assessment based on the current model or system of beliefs. When strong evidence arises against the adequacy of the current theory or model, the next Creation step is called for.
\end{description}




Given the creation and evaluation steps, the \articulation~step is seemingly redundant but can be viewed as to correspond to what Kuhn \citep{kuhn1970structure} calls `normal' research. While philosophically similar to Whewell's structure of scientific discovery and drawing upon Kuhn's theory of scientific revolutions \citep{kuhn1970structure}, TB is formulated in a principled manner with a sound logic of science to understand and analyze scientific discovery. Consequently, TB is quantifiable, as demonstrated in Section \ref{s:mnms}, showcasing the potential of TB for strong AI. It can also be used to evaluate scientific methods and theories, even including inferential theory itself, as shown in Section \ref{s:logic-of-science}. These examples are intentionally chosen for simplicity and conceptual clarity for general readers.

\section{A Simple Illustration: Many Normal Means}\label{ss:mnms}\label{s:mnms}

Scientific discovery involves formulating and, when necessary, refining or rejecting hypotheses about the nature of the world. As we acquire evidence, we create models that we believe capture the structural mechanisms giving rise to our data. Over time, we may refine our parameter estimates for these models, and slowly the models will both fit our existing data better, and have higher predictive capacity. However, we will occasionally encounter situations where new data appear to entirely discredit an existing model. Falling victim to the sunk cost fallacy, these new data may initially be rejected on the basis of their nonconformity, and scientific progress may stall due to a hesitation to explore alternative theories when so much work has already been devoted to determining implications of the existing model. Nevertheless, statistical tools are available for formulating and testing whether an old model should be rejected in light of new data. This is particularly pertinent today. In the modern era of deep learning, foundation models are trained on massive datasets and then fine-tuned for specific use cases, but it is not often clear when a model should be fine-tuned as opposed to outright retrained. A special case, which we explore for the purpose of illustration, is the problem of many normal means, where we ask whether a normal mixture model with a given number of components and previously accepted set of means can reasonably concord with a new observation. This situation arises in repeated experiments, where the question becomes whether we are observing a more diverse population that previously thought. For example, if you were to take note of birds spotted in a prairie and you had four categories of known birds, you may at first mistakenly classify a new, but similar, bird into one of the existing categories. However, as measurement accuracy or sample size increases, it may become clear that the previous four categories are insufficient to describe the population.

More formally, the many-normal-means problem 
is about making inference
on the unknown means $\mu_1, ..., \mu_n$ from the sample $Y_1,..., Y_n$
with the model
\[
	\mu_i \sim {\mathbb P}_\theta\quad\mbox{ and }\quad
	Y_i|\mu_i \sim N(\mu_i, 1)\qquad(Y_i\in\mathbb{Y}, \ \theta \in \Theta^{(n)} \subseteq \mathbb{R}^p)
\]
for $i=1,...,n$ \citep[see, {\it e.g.}][]{efron_empirical_2016,jiang2024estimation}. The inferential goal is to 
estimate $\mu_i$, to model $ Y_i$, to predict a new observation, or to produce uncertainty quantification on assertions of interest. As a further simplification of this problem for our illustration purpose, suppose that $\mathbb{P}_\theta$ is a normal mixture model with an unknown number of components, $K$, each with the same known, small variance, $\sigma^2$. To be clear, models with different $K$ values here are meant to be different models in our illustration. Therefore, we have \begin{equation}
\mu_i \sim \sum_{k=1}^K \pi_k N(\phi_k, \sigma^2), \qquad \sum_{k=1}^K \pi_k = 1. \label{eq:normal-mixture-setup}
\end{equation} 
Using our samples, $Y_i$, we aim to perform inference on $K$, $\pi_1, \dots, \pi_K$, and $\phi_1, \dots, \phi_K$ in an iterative sense. Let $g_{n,K}(Y_1, \ldots, Y_n): \mathbb{R}^{n} \rightarrow \mathbb{R}^{2K}$ be an estimator for $\theta_K \equiv $ 
$(\pi_1, \dots, \pi_K,$ $ \phi_1, \dots, \phi_K)$ $\in \mathbb{R}^{2K}$, and let $h_n(Y_1,\ldots,Y_n) : \mathbb{R}^n \rightarrow \mathbb{Z}^+$ be an estimator for $K$. We refer to $h_n$ as the transformative level and $g_{n,K}$ as the exploratory level of the model. An iterative procedure would utilize the transformative estimator $h_{n-1}(Y_{-k}) = K_{n-1}$ and its corresponding exploratory estimate $g_{n-1, K_{n-1}}$ along with the new sample, $Y_k$, to \textit{evaluate} hypotheses articulated in the \articulation~step.

Beyond this one initial sample, the process of gathering new observations creates a time series which requires continual evaluation and, sometimes, creation or \articulation. In the case above, we had a single new observation, and we explored our landscape of conclusions under that new observation. Now, suppose that we have a set of observations 
\[\mbox{$\{Y_t:\; Y_t \in \mathbb{R};\; t \in \mathbb{Z}^+\} $}\]
where $Y_{1:n} = Y_1, \dots, Y_n$ are initial observations and $Y_{n+1}, Y_{n+2}, \dots $ are future observations. Beginning at time $n$, we must create (creation) an initial model and estimate (articulate) the parameters for that model. Then, at each point $t \in [n+1, \infty)$, we perform an evaluation step to see if the current model reasonably explains the new data. If that evaluation step leads us to reject the model, then we must continue with a new creation and \articulation~procedure. Next, we outline the details of each of these steps, continuing with the setup in \eqref{eq:normal-mixture-setup}. 

It should be noted that for our exposition, familiar simple statistical procedures are used. Alternative methods such as those based on inferential models (see Section \ref{s:logic-of-science}) can be considered, especially in future research in developing TB creativity systems.

\subsection{Creation}

\begin{figure}
 \begin{centering}

    \includegraphics[width=0.45\textwidth]{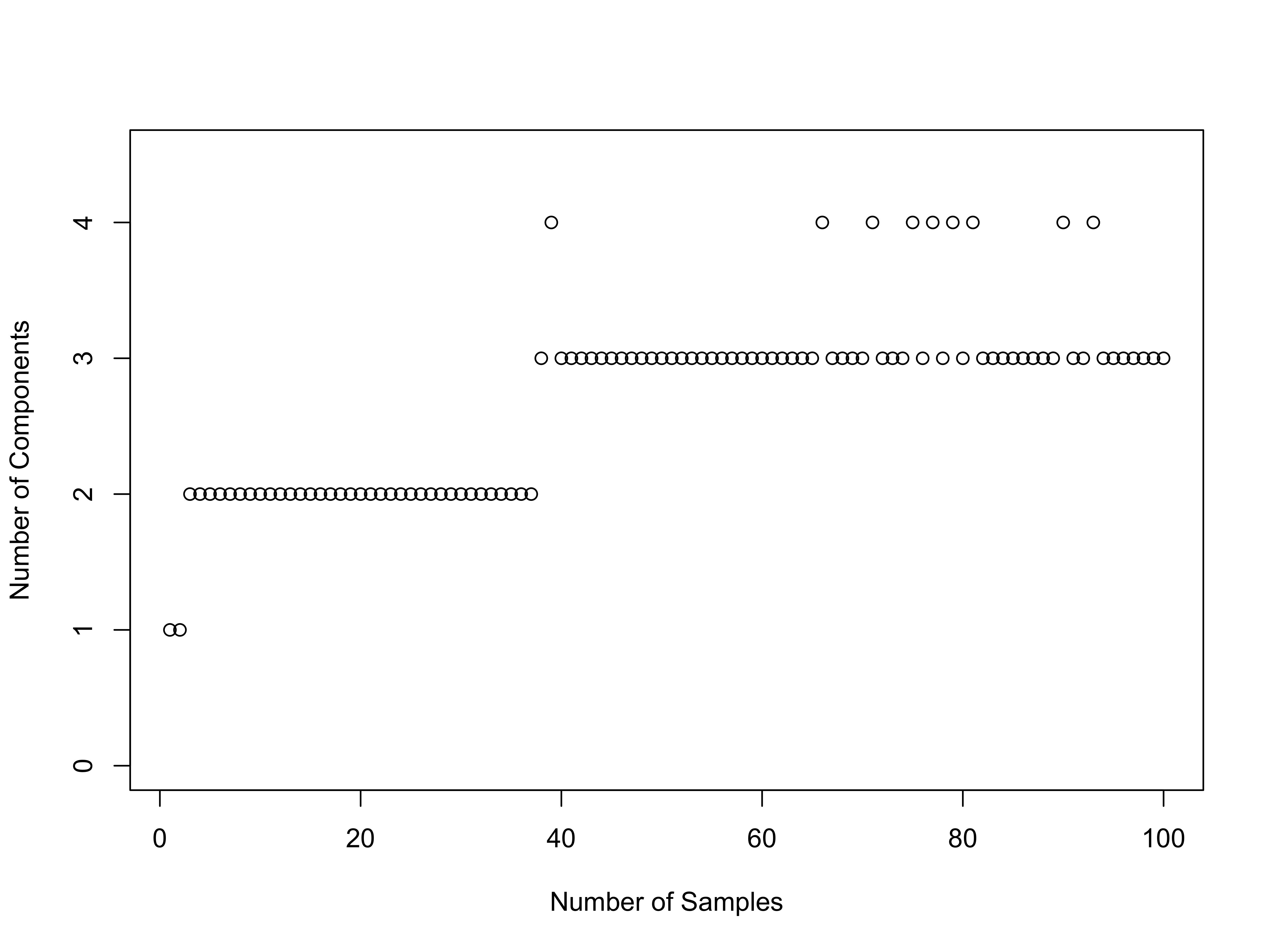}
    \includegraphics[width=0.45\textwidth]{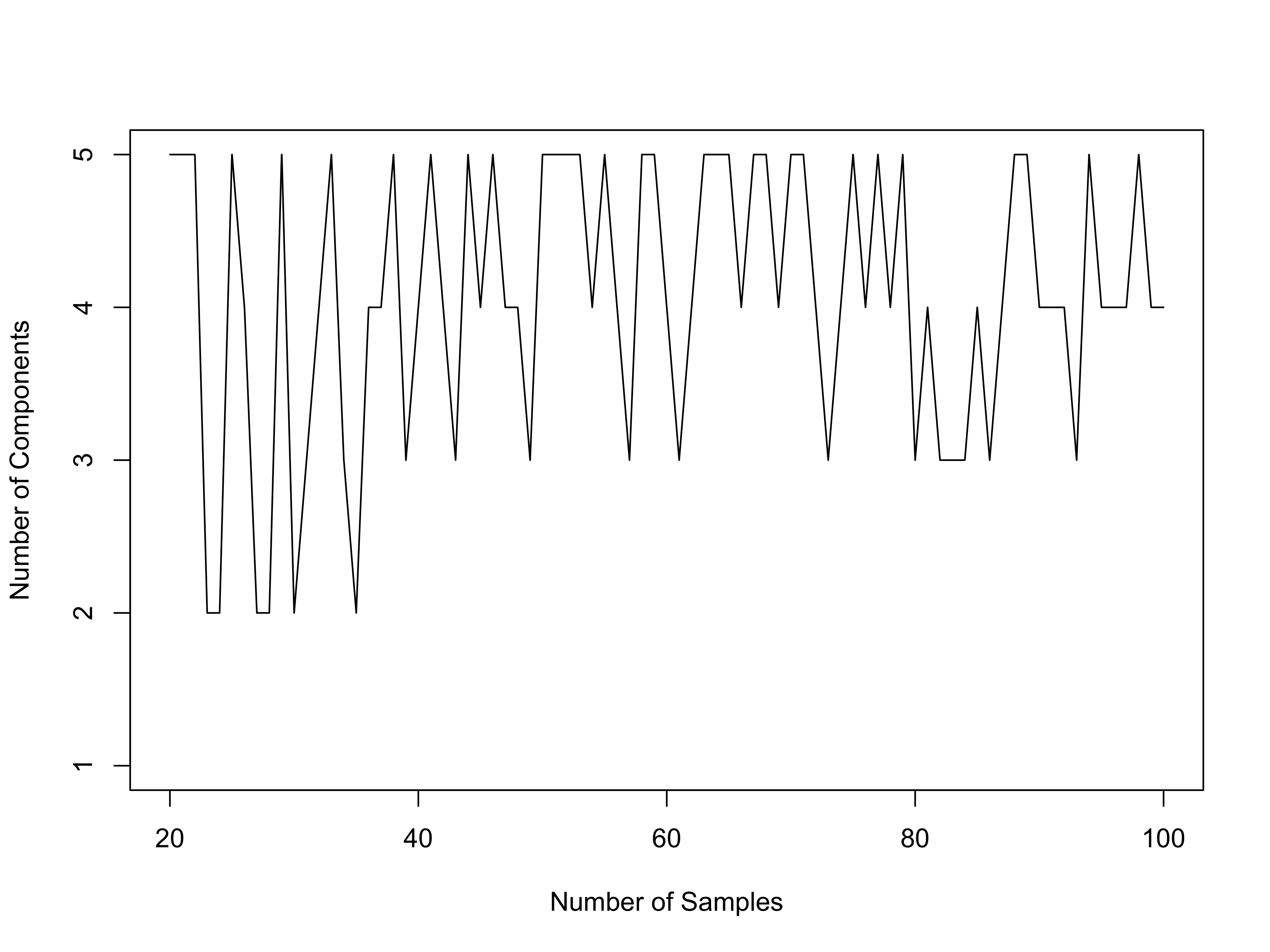}
     \caption{Initial estimates of the number of components by sample size using BIC (left) or 80-20 cross validation of the negative log likelihood (right) when the true model has $K=3$, $\theta = (-2\quad 2 \quad 5)^\top$ and $\pi = (0.3\quad 0.5\quad 0.2) ^\top$.}
      \label{fig:creation-mnm}
 \end{centering}
 \end{figure}

In the creation step, we calculate an initial estimate of $K_n$. This estimate does not need to be optimal, because we can test against other hypotheses in the \articulation~step, but it should be reasonable in the sense that the model fits the data. 
One way to estimate the number of components is using penalized maximum likelihood, such as BIC \citep{schwarz_estimating_1978} where we optimize
$$\text{BIC} = -2 \ell(\theta | Y_1,\dots, Y_n) + K\log(n) = C(\theta)-2\sum_{i=1}^n \log \sum_{j=1}^K \pi_j\exp\left\{-\frac{(Y_i-\pi_j)^2}{2}\right\} + K\log(n)$$
 by first minimizing $-2\ell(\theta|Y_1,\dots,Y_n)$ for each $K \in 1,\dots,n$ separately and using the $K$ which produces the lowest BIC. By generating a sequence $Y_1, Y_2,\dots$, we can see how the initial estimate of $K$ changes based off how many samples we have. A demonstration of this is shown on the left hand side of \autoref{fig:creation-mnm}. We note that while most of the time, a larger sample provides a more accurate estimate for the true $K$, sometimes this is not the case, even for fairly large sample sizes. Now, if instead of using BIC, we used cross validation with $50$ replications of an $80-20$ split of training and testing, and then chose the number of components with the lowest mean negative log likelihood for each sample size, the behavior is even more oscillatory between various values of $K$, as seen on the right hand side of \autoref{fig:creation-mnm}. Both of these methods, however, would provide decent initial estimates for $K$ that are 
 useful.

To estimate $\pi_1, \dots, \pi_K$ and $\phi_1, \dots, \phi_K$, two common options are maximum likelihood using the Expectation-Maximization (EM) algorithm or sampling from a Bayesian posterior with a prior distribution $$\pi \sim \text{Dirichlet}(\alpha_1, \dots, \alpha_K) \qquad \alpha_1 = \alpha_2 \dots = \alpha_K = 1.$$
For our numerical experiments, we implemented a standard Gibbs sampler and the EM algorithm (see \Supplementary~B).

\subsection{\Articulation~and Evaluation}

The evaluation step in scientific creativity is comprised of hypothesis testing and falsification. Given existing or new data, we check whether the model created during the creation step reasonable explains the data  in the context of the alternatives articulated above. In this case, we articulate the hypotheses $H_0: K_{n-1} = K_n$ versus $H_a: K_{n-1} \neq K_n$. If we fail to reject $H_0$, then a new corresponding exploratory estimator $g_{n, K_{n-1}} (Y_1, \dots, Y_n)$ can be calculated. Alternatively, if we reject $H_0$, we estimate $K_n = h_k(Y_1, \dots, Y_n)$ and $g_{n, K_{n}}$, and we say that $Y_k$ catalyzed a transformative discovery. 

To take the simplest case, consider that we have a sample $Y_1, \dots, Y_{n-1}$, and that $h_{n-1}(Y_{-n}) = K_{n-1} = 1$. This implies that 
$\pi_1 = 1$, and a reasonable estimator for $\phi_1$ is the sample mean, $\bar Y_{-n}$. Now, we observe $Y_n$ and test whether $H_0: K_n = K_{n-1} = 1$ or $H_a: K_n = 2$. Let $0<\alpha<1$ be our confidence level, and we reject $H_0$ if 
$1-\Phi\left(\left|\frac{\bar Y_{n-1} - Y_n}{\sqrt{\sigma^2 + 1}}\right|\right)< \frac{\alpha}{n}$, where $\Phi(.)$ denotes the cumulative distribution (\textit{c.d.f.}) function of the standard normal distribution.
Here we use the Bonferroni adjustment for multiple testing, because each of the $n$ samples is assumed to be {\it i.i.d.}~under the model for $Y_{-n}$. If we fail to reject $H_0$, then we can update our mean estimate as $\phi_1 = \frac{(n-1)\bar Y_{n-1} + Y_n}{n}$. However, if we reject $H_0$, then we need to estimate all of $\pi_1, \phi_1,$ and $\phi_2$. Note that because $\pi_2 = 1-\pi_1$, we need not estimate $\pi_2$ directly. Therefore, we are fitting the model
$$f(x; \phi_1, \phi_2, \pi_1) = \frac{1}{\sqrt{2\pi}}\left[\pi_1\exp\left(-\frac{(x-\phi_1)^2}{2}\right) + (1-\pi_1)\exp\left(-\frac{(x-\phi_2)^2}{2}\right)\right],$$
which can be done by maximum likelihood using the EM algorithm as described in \Supplementary~B.

\begin{figure}
    \begin{centering}

        \includegraphics[width=\textwidth]{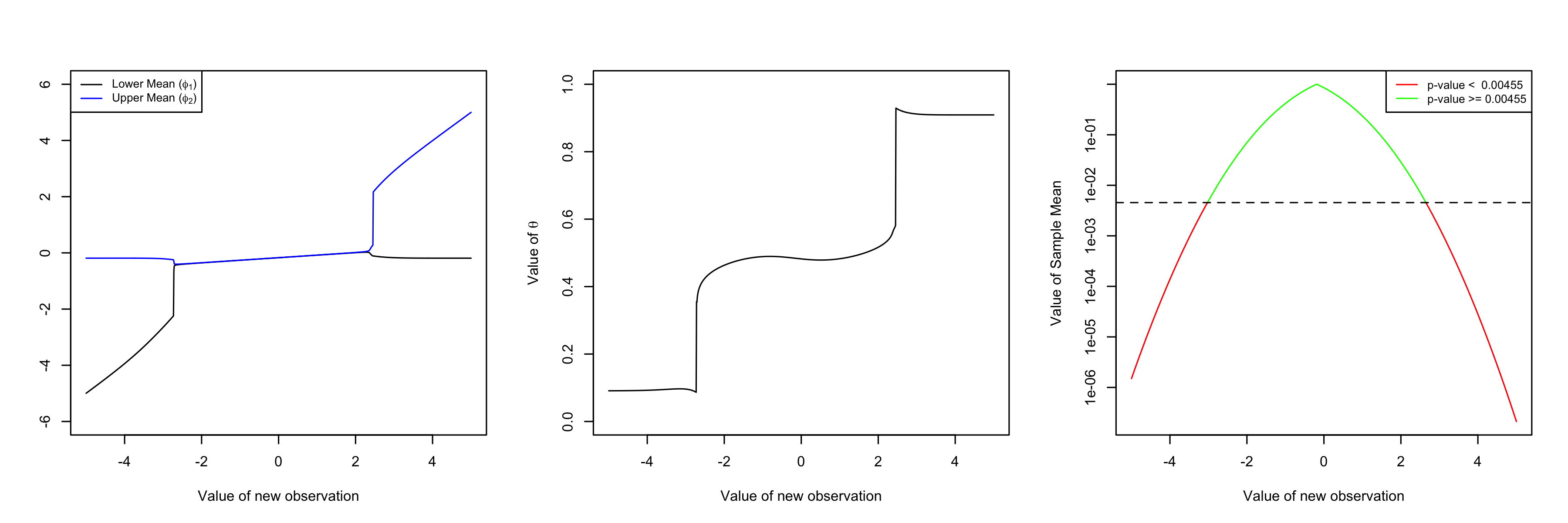}
        \caption{An illustration of the effect of one new observation on model specification for an initial sample with $K_n = 1$, $X_i \sim N(0,1)$, and $i \in 1,\dots,n=10$.}
        \label{fig:two-means-em}
        
    \end{centering}
\end{figure}

Now, we assume that $n=10$ and that we previously concluded that our data follow a standard normal centered at their sample mean. That is, $K_{10} = 1$, $\pi_1 = 1$, and $\phi_1 = \bar{Y}$. We analyze what occurs when we observe a new value, $Y_{11} \in (-5,5)$, and re-estimate the model at both the exploratory and transformative levels. The right-most plot of \autoref{fig:two-means-em} shows the exploratory (green) and transformative (red) regions of observations for $Y_{11}$. If $X_{11}$ falls within the exploratory region, then the model is refined, rather than rejected. However, if the model is rejected (red), then we have additional parameters to estimate. The left two plots show the behavior of the MLE under the new observation given that we reject the current model. These plots showcase a region of stability of the MLE with both estimated means around $0$ when the new observation falls close within the range of previously observed values. The shift is reflected in the middle plot as well, which indicates that the model essentially merges from two components back to one when the new observation is indistinguishable in distribution from existing values. In scientific inquiry, analyses such as these are important, because they provide a test-bed for falsifiability. We are asking what data would lead us to reject our current model, and under what reformulating our model would entail given various new observation values.

\ifthenelse{1=1}{}{
\subsection{The discovery of Neptune: a missing-data example
}\label{ss:neptune-revisited}
\begin{comment}[TODO]\color{magenta}
A problem is that the proposed two illustrative examples are not new.
   Do we have any new applications to demonstarte the advantage of our method? 
   The discovery of Neptune may be a good example.
   One way to consider this problem is  to view it as a missing data problem. The generality of missing data problem makes it more interesting (?).
\end{comment}
}







\section{The 260-Year Quest for a Unified Logic of Science} 
\label{ss:somegpt}\label{s:logic-of-science}

There have been too many creative works in statistics that would require a book-long space to discussion \citep[see, {\it e.g.},][]{stigler2002statistics,bickel2015mathematical,gelman2021most}. 
In this section, we take a look at the quest for a unified logic of science, the most unsolved problem in statistics \citep{efron2012bayesian},  from a \weak~beliefs perspective. 

In \cite{gigerenzer2015cognition}, the authors argued that in the science of mind, theories are particularly likely to come from tools, and they are specially concerned with the emergence of the metaphor of the mind as an intuitive statistician. Here, tools, physical for changing environment or logical for reasoning, are certainly products of creativity. Their observations are particularly intriguing in the context of strong AI research. 
These include their recognized two scientific revolutions, cognitive and probabilistic, in the middle of the twentieth century. Our example here focuses on the latter and its more general form --- statistical. Indeed, all is about reasoning with uncertainty and has a long history \citep{nickerson2004cognition}. Our brief discussion will focus on inventions for inductive inference that date back to the 260-year-old topic --- \cite{bayes1763essay}.

\ifthenelse{1=1}{

}{
\addText{\color{magenta}Read \cite{gigerenzer2015cognition} on the discussion on the transition from tools to theory. To science, of course, theories are tools. TODO:
\begin{itemize}
    \item Summarize and comment on  \cite{gigerenzer2015cognition} and \cite{nickerson2004cognition}; 
    \item Review the critiques of Cognition as Intuitive Statistics; and
    \item Comment on the modern AI and cognition, from a statistical perspective.
\end{itemize}
}

\newcommand{\minp}[1]{\begin{minipage}{1.5in}#1\end{minipage}}
\newcommand{\mine}[1]{\begin{minipage}{1.5in}#1\end{minipage}}
\newcommand{\mina}[1]{\begin{minipage}{1.0in}#1\end{minipage}}
\newcommand{\minc}[1]{\begin{minipage}{2.5in}#1\end{minipage}}

\begin{table}\color{blue}\spacingset{0.82}\small
\begin{tabular}{cclclcl}
\multicolumn{7}{l}{\color{red}\it Here is a revised data table focusing on the historical and creative events in the foundations of }\\
\multicolumn{7}{l}{\color{red}\it statistical inference, emphasizing conceptual advances, contributors, and their significance:}\\
\multicolumn{7}{c}{\color{red}\bf FIX THE DATA $\checkmark$}\\
\hline\hline
    Year &&	Event / Contribution	&& Authors	&& Significance / Notes\\
    \cline{1-1}\cline{3-3}\cline{5-5}\cline{7-7}
    \multicolumn{7}{c}{}\\
\multicolumn{7}{c}{}\\
\textbf{1763}	&&\minp{\color{magenta}  Bayesian Inference}	&&\mina{Thomas Bayes}	&&\minc{Laid the groundwork for Bayesian probability, introducing the idea of updating beliefs with data.}\\
\multicolumn{7}{c}{}\\
\textbf{1930}	&&\minp{\color{magenta} Fiducial Inference	}	&&\mina{R.A. Fisher	}	&&\minc{Proposed an alternative to Bayesian and frequentist methods, sparking debates on inference logic.}\\
\multicolumn{7}{c}{}\\
1960s && \minp{\color{magenta}
Belief Functions	}	&&\mina{Arthur P. Dempster (1967), Glenn Shafer (1976)}	&&\minc{	This theory generalized Bayesian probability by introducing belief and plausibility measures, allowing for more flexible modeling of uncertainty in cases with incomplete or imprecise information. It has been widely applied in fields like artificial intelligence, expert systems, and sensor fusion.}\\
\multicolumn{7}{c}{}\\
2000s	&&\minp{\color{magenta}Inferential Models}	&&\mina{Ryan Martin, Chuanhai Liu	}	&&\minc{Bridged frequentist and Bayesian approaches, emphasizing validity and efficiency in inference.}\\    
\multicolumn{7}{c}{}\\

\hline
\end{tabular}
\caption{[FIX THE TABLE CONTENT] Creativity observations, consisting of historical and creative events, in the development of the bootstrap resampling method.}
\label{tbl:inference}
\end{table}

\addText{The last century has seen tremendous efforts on the development of statistic inference, including those events summarized in Tables \ref{tbl:inference} and \ref{tbl:inference-expanded}.
Our focus will be only on sensible answers to the question: What would be a desirable theory of inference for strong AI? We assume that tools for gaining knowledge from experience as the most fundamental theory in intellectual activities. We consider those tackling inference problems as a general theory that can be viewed as the logic of science.} This classification is difficult because many frequentist procedures can be converted for such a purpose, especially from the IM perspective, such as Fisher's P-value \cite{martin2014note}.
Table \ref{tbl:inference} underscores pivotal developments in the foundations of statistical inference, reflecting a...  
}




\subsection{The Bayesian invention}

There have been debates on whether scientific tools such as logic, mathematics, and statistics belong to science. This is particularly relevant because the proposed TB framework we consider here is based on empirical science. Our simple solution is to
consider the content for which the statistical method is designed. To be specific, consider the scientific problem of \cite{bayes1763essay} from an experiment consisting of $n$ independent and identically distributed Bernoulli trials with the probability of success $\theta \in [0,1]$. From this experiment, we have observed $X$ successes, $X\leq n$. We want to update our knowledge on $\theta$ or the probability calculations of future such Bernoulli events, assuming the unknown probability of success $\theta$ follows the standard uniform distribution $\mbox{Uniform}(0,1)$ {\it a priori}. 
The result is the Bayes theorem: given $X$, $\theta$ has the posterior distribution
\[
\pi(\theta|X) = \frac{\pi(\theta)f_\theta(X)}{\int_0^1 \pi(\theta)f_\theta(X) d\theta},
\]
where $\pi(\theta)$ is the probability density function of the prior distribution $\mbox{Uniform}(0,1)$, $\pi(\theta)=1$ for all $\theta \in [0,1]$, and $f_\theta(X)$ is the probability mass function of $X$, $\theta^X(1-\theta)^{n-X}.$ 

Bayes' theorem and Bayesian inference developed over the last 260 years, including the work of Pierr-Simon Laplace \citep[see, {\it e.g.}][]{jaynes_probability_2003,stigler1986laplace} and neo-Bayesian revival \citep[see, {\it e.g.}][]{fienberg2006did}, have profoundly shaped mathematics, statistics, and numerous other disciplines and, thereby,  gone through a careful process of {\it TB-\articulation} since their inception or {\it creation}. 
Nevertheless, while Bayesian remains as a popular school of thought, the dominant school of thought in scientific inference is frequentist, which started to surface a century ago. Here, we take a critical look the Bayesian theory from a perspective of {TB-evaluation}.


Perhaps, the most acceptable evaluation scheme is to evaluate the predicted future observations from the Bayes model against the outcome that will be actually observed. Mathematically, this ultimately reduces to the {\it frequentist evaluation} of the posterior distribution when a specific prior distribution $\pi(\theta)$ is used, even though the data analyst has no prior knowledge about the value of $\theta$ {\it a priori}. The failure of the Bayes theory on frequency evaluation (TB-evaluation), especially on constrained parameter and multi-parameter problems, inspired paradigm shift research for alternative ways of scientific inference. 

\ifthenelse{1=1}{}{
Additionally, Bayesian probability theory has also been intended to serve as the {\it logic of science}. The other evaluation would be on this aspect and would lead to scrutinization of Bayes factors that have used for hypothesis testing. Historically, this has not appeared to be under the spotlight of debates, compared to the need for a frequency evaluation. Part of the reason could be that the possible alternatives, frequentist methods of hypothesis testing, are themselves questionable to form a theory as the logic of science. This issue is discussed further in Section \ref{ss:logic}.

A comprehensive review of Bayesian theory is given by \cite{fienberg2006did}. Notably, \cite{fienberg2006did} wrote:
`{\it Laplace's introduction of the notion of ``indifference'' as an argument in specifying
a prior distribution was first in a long line of efforts to discover the statistical holy
grail: prior distributions reflecting ignorance. The search for the holy grail continues
today under a variety of names, including objective Bayes, and in response to every
method that is put forward we see papers about logical and statistical inconsistencies
or improprieties. ... Mathematicians and philosophers continued to debate the meaning of probability throughout the remainder of the nineteenth century. For example, Boole, Venn, Chrystal, and others argued that the inverse method was incompatible with objective probabilities, and they also critiqued the inverse method because of its use of indifference prior distributions. 
But when it came to the practical application of statistical ideas for
inferential purposes, inverse probability ruled the day (c.f. Zabell (181; 182)).14 This
was largely because no one came forward with a systematic and implementable ``frequentist'' alternative. \footnote{Thus, in retrospect, it shouldn't be surprising to see inverse probability as the method of choice of the great English statisticians of the turn of the century, such as Edgeworth and Pearson. For example, Edgeworth (49) gave one of the earliest derivations of what we now know as Student’s t-distribution, the posterior distribution of the
mean µ of a normal distribution given uniform prior distributions on $\mu$ and $h = \sigma^{-1},$ calling it the ``subexponential'' distribution. Of
course, today we also know that the result was derived even earlier, by L\"{u}roth in 1876,
who also used inverse probability methods (see Pfanzagl and Sheynin (125)).}}'
}

\subsection{The fiducial inspiration}\label{ss:fiducial}

Philosophically, the primary foundation of frequentism lies in the interpretation of probability as a measure of long-run frequency of events in repeated trials. This perspective was principally developed (created) by Ronald Fisher, as well as Jerzy Neyman and Egon Pearson. Ronald Fisher made significant contributions by introducing the concept of significance testing, which evaluates how surprising a statistic is with respect to a null hypothesis.

Neyman and Pearson expanded on Fisher's ideas to address scenarios involving multiple competing hypotheses. They proposed that the likelihood ratio, comparing probabilities under different hypotheses, could be used to maximize the differentiation between these hypotheses. Their work introduced the formal framework of Type I and Type II errors, with Type I errors representing false positives (rejecting a true null hypothesis) and Type II errors representing false negatives (failing to reject a false null hypothesis). This framework also established the concept of test power, emphasizing the optimization of tests to balance error probabilities while exceeding a predetermined significance level.
\ifthenelse{1=1}{}{
It may be helpful to have a good understanding of the history leading to  frequentist statistics, starting from what \cite{fienberg2006did} wrote:
`{\it At the time that Ronald Alymer Fisher began his studies of statistics at Cambridge in
1909, inverse probability was an integral part of the subject he learned (c.f. Edwards
(50)). Frequentist and other non-Bayesian ideas were clearly “in the air,” but it is
diﬃcult to know to what extent Fisher was aware of them. For example, as a student
he might have been led to read papers and books dealing with such alternatives by
his teacher, F.J.M. Stratton, but he rarely cited relevant precursors or alternatives to
his own work (see Aldrich’s (1) account of what Fisher studied). Over a period of 10
years, from 1912 to 1922, Fisher moved away from the inverse methods and towards
his own approach to inference he called the ``likelihood,'' a concept he claimed was
distinct from probability.}'   
\addText{According to {\it Wikipedia} with the reference on ``Sir Ronald Fisher'' by Russell, E. John Russell,}  
R. A. Fisher worked at the Rothamsted Experimental Station from 1919 for 14 years; 
there, he analyzed its immense body of data from crop experiments since the 1840s, and developed the analysis of variance (ANOVA). He established his reputation there in the following years as a biostatistician. It would be interesting to know how that applied statistics experience helped shape Fisher's frequentism. 

The Neyman-Pearson school of thought certainly is revolutionary and dominates applications of statistical inference. The historical account on the difference between the Neyman-Pearson school and Fisherian school of thought on inference is clear from what \cite{fienberg2006did} wrote:
\begin{quote}
Fisher’s work had a profound inﬂuence on two other young statisticians working in
Pearson’s laboratory at University College London: Jerzy Neyman, a Polish statistician
whose early work focused on experiments and sample surveys, and Egon Pearson, Karl
Pearson’s son. They found Fisher’s ideas on signiﬁcance tests lacking in mathematical
detail and, together, they set out to extend and “complete” what he had done. In the
process, they developed the methods of hypothesis testing and conﬁdence intervals that
were to revolutionize both the theory and the application of statistics. Although Fisher
often disagreed with them caustically and vigorously both orally and in print (e.g.,
see Fienberg and Tanur (61)), some amalgam of their approaches—referred to later as
“frequentist” methods—soon supplanted inverse probability. The mathematical foun-
dation for these methods emerged from Russia and the axiomatic work of Kolmogorov
(93). The methods quickly spread to diverse areas of application, for good and for bad.
Gigerenzer et al. (69) describe the impact of these frequentist methods especially in
psychology. 
\end{quote}
}
Nevertheless, from the TB perspective of refining Bayesian theory that has to be probabilistic to serve as the logic of science, Fisher's concept of significance testing using his invention of p-values is aligned with the logic of science, and his fiducial inference seemed to be on target more than Neyman-Pearson's concepts of confidence of intervals and their framework of Type I and Type II for hypothesis testing. The retrospective discussion of Fisher's p-value in the IM framework is given by \cite{martin2014note}. Here, we discuss fiducial inference with the focus on our TB analysis using the example of \cite{bayes1763essay}, rather than the simpler, more familiar cases of \cite{fisher1973statistical}.


\begin{figure}[!htb]
\begin{center}
\includegraphics[width=0.70\linewidth]{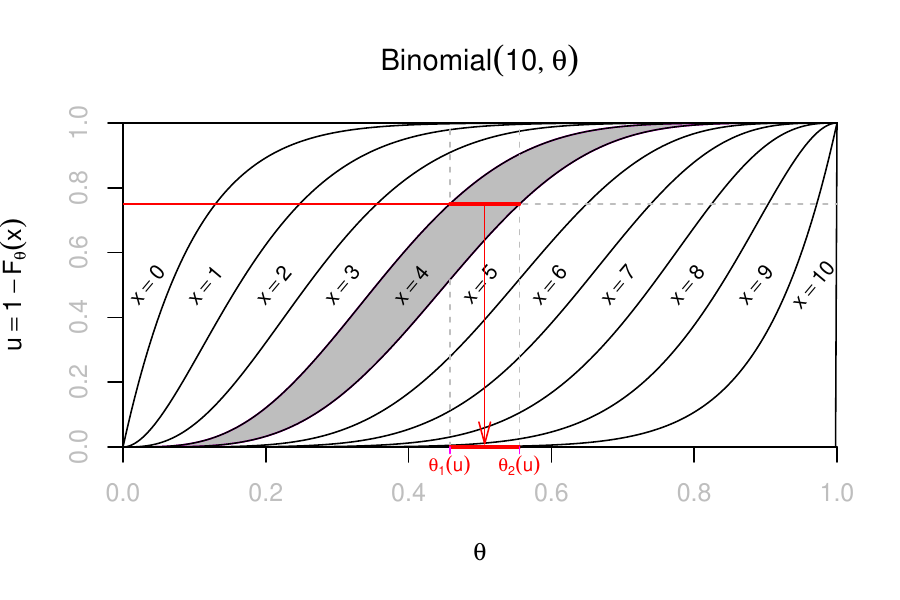}
\vspace{-0.15in}
\caption{The {\it fiducial} set-valued mapping $\{\theta: F_\theta(x-1) < 1-u \leq F_\theta(x)\}$ for $u$ given $x$. The gray area is for the case with $n=10$ and $x=4$.
}
\label{fig:binomial-ims}
\end{center}
\end{figure}

Since R. A. Fisher didn't develop a complete fiducial theory, fiducial-inspired efforts have appeared in different places \citep[see, {\it e.g,}][]{zabell1992ra,hannig2009generalized}. 
The $\mbox{Binomial}(n, \theta)$ model for the observed count $X$ of successes in $n$ iid Bernoulli trials with the probability of success $\theta$ has the probability density function (p.d.f.)
\begin{equation}\label{eq:binom-f}
   f_\theta(x) = {n \choose x} \theta^x (1-\theta)^{n-x}
   ,\quad(k=0,1,2,..., n)
\end{equation}
and the c.d.f.
\(
F_\theta(x) 
   = 1 - \mbox{Beta}_{x+1, n-x}(\theta),
\)
the regularized incomplete beta function
$I_{1-\theta}(n-x, x+1)$.
Note $\mbox{Beta}_{a, b}(x) = I_x(a, b)$,
$I_x(a, b) = \int_0^x t^{a-1}(1-t)^{b-1}dt/
\int_0^1 t^{a-1}(1-t)^{b-1}dt$, and
$I_x(a, b) = 1- I_{1-x}(b,a)$.
Let $U \sim \mbox{Uniform}(0,1)$, and define $X$
to be the $U$-th quantile of $F_\theta$,
\[
  X = F_\theta^{-1}(U) = \min\{x: U \leq F_\theta(x)\},
\]
which serves as a data generating equation with $U$ as what \cite{martin2013inferential} call {\it the auxiliary variable}.
Then, we have for $U=u$ and $x = F_\theta^{-1}(u)$
\begin{equation}\label{eq:binomial-ims}
 \mbox{Beta}^{-1}_{x, n-x+1}(1-u)
 < \theta \leq \mbox{Beta}^{-1}_{x+1, n-x}(1-u).
\end{equation}
This {\it fiducial} set-valued mapping, is depicted in Figure \ref{fig:binomial-ims}.
For continuous distributions, This {\it fiducial} set-valued mapping reduces to a usual function. In that case, as shown in \cite{fisher1973statistical}, the fiducial distribution of $\theta$ is obtained by pushing the distribution of $U$ to the space of $\theta$ using this usual function.
The discussion of the general set-valued mapping will be continued in Section \ref{ss:DS}.

Unfortunately, R. A. Fisher did not give a general definition of fiducial inference. Most of his examples were for a single parameter, except for the famous Behrens-Fisher problem \citep[see, {\it e.g.},][]{martin2015marginal}; different generalizations have been given when there are several parameters \citep{tukey1957some,stein1959example}. 
According to Zabell (1992), ``the ﬁducial argument stands as Fisher’s one great failure'', a sentiment that has been echoed by others \citep{dawid2024fiducial}, and apparently has a connection with its perception by Jerzy Neyman, one of Fisher's contemporaries and critics. Neyman and Fisher were known to have significant philosophical and methodological disagreements, particularly regarding statistical inference. Neyman, a key proponent of frequentist methods, criticized Fisher's fiducial inference approach as being vague and inconsistent with strict frequentist principles. All these can be viewed as failures with respect to TB-evaluation in terms of frequency evaluation.

R. A. Fisher himself maintained throughout his life that fiducial inference was an important contribution, even if it was not universally accepted or fully developed to his satisfaction. Critics like Neyman often emphasized its logical difficulties and the lack of general applicability, leading to the sentiment that it was a notable misstep in Fisher's otherwise groundbreaking career.

Nevertheless, Fisher's idea of fiducial inference has appeared to be inspirational.  
It stimulated the tremendous explorations \citep[see, {\it e.g.}][and references therein]{birnbaum1961confidence,fraser1961fiducial,dempster1964difficulties,zabell1992ra,dawid2024fiducial,nancy2022fraser},
and continues to inspire creative statistical methods. The latter includes generalized 
fiducial \citep{hannig2009generalized,hannig2016generalized,liang2024extended}, which are mostly focused on large-sample-based justification, applications, computational methods, and exact Neyman-Pearson confidence methods \citep{cook1990confidence,xie2013confidence,cui2022demystifying,xie2022repro}.





\subsection{The Dempster-Shafer discovery}\label{ss:DS}

As (the creation of) both a successor fiducial and a generalization of Bayes (
see \cite{dempster2008dempster}), the Dempster-Shafer theory builds upon Dempster's discovery of the need of using upper and lower probabilities for inference and Shafer's development of a broader framework \cite{shafer1976mathematical} to define and manipulate belief functions. Its dual notions of belief and plausibility offer a nuanced way to assess confidence in propositions when full probabilistic information is unavailable. 

Here, from the TB perspective, consider the running example of inference with a binomial count. A natural way of extending Fisher's fiducial approach to continuous examples to the discrete cases would lead to considering the set-valued mapping \eqref{eq:binomial-ims}:
\begin{equation}\label{eq:DS-set-value}
\Theta_{x}(u) \equiv \{\theta: \; \mbox{Beta}^{-1}_{x, n-x+1}(1-u)
 < \theta \leq \mbox{Beta}^{-1}_{x+1, n-x}(1-u)\}.
\end{equation}
Using a predictive interval for the unobserved auxiliary variable $U$, for example,
\[\mathcal{U}_{\alpha} = \left[\frac{\alpha}{2}, 1-\frac{\alpha}{2}\right],\qquad \alpha \in(0,1),\]
we can construct a $(1-\alpha)100\%$ frequentist confidence interval
\[
\Theta_{x}(\alpha) = \cup_{u\in \mathcal{U}_{\alpha}}\Theta_{x}(u) = \left[\mbox{Beta}^{-1}_{x, n-x+1}(\alpha/2),\; \mbox{Beta}^{-1}_{x+1, n-x}(1-\alpha/2)\right]
\]
for $\theta$, which corresponds to the method of \cite{pearson1920fundamental}. Interestingly, this supports the perception that frequentist ideas are ``in the air'' when R. A. Fisher became what we now consider as frequentist. 

In addition to the discovery of the necessity of using set-valued inverse mapping and, thereby, lower-and-upper or imprecise probabilities, other innovations of the Dempster-Shafer theory include combining information, Dempster's rule of combination, and the mathematical theory of evidence \citep{shafer1976mathematical}. From the TB perspective, everything seems to have come together to form a satisfactory logic of science, except for the frequency evaluation that the majority of scientists apparently considered important logically by the nature of science. Viewed as TB-evaluation, this is discussed in depth in \cite{martin2010dempster} and \cite{zhang2011dempster}, which eventually lead to the work discussed next in Section \ref{ss:ims}. For an extensive review of the Dempster-Shafer theory and more discussion of its frequency properties, see \cite{yager2008classic}, \cite{liu2015frameworks}, \cite{denzux201640}, and \cite{denoeux2018frequency}.




\subsection{The inferential models framework}\label{ss:ims}

Creating a fully satisfactory logic of science is probably still an unsolved problem. This was evidently in statistics and science near the turn of this millennium. 
For example, John W. Tukey (1990) said:
``{\it
Today I do not believe there is any chance for a single unifying approach to inference.}''
and Bradley Efron (2013) wrote:
``{\it
... perhaps the most important unresolved problem in statistical inference is the use of Bayes' theorem in the absence of prior information.}''
But, there is an encouraging assessment by a forthcoming review of IMs in {\it Statistical Science} \citep{cui2022demystifying}. Those authors say:
\begin{quote}
\textit{
IMs brought a thoroughly novel idea into the foundations of statistics by formalizing a way to assign epistemic probabilities to events that have guaranteed frequentist interpretation. ...  They provide a powerful argument for anyone seeking fiducial or objective Bayes distributions on parameter space to consider making calculations on the auxiliary [variables].}
\end{quote}
Here, we take a critical look at IMs from the perspective of TB using the running binomial example.

\begin{figure}[!htb]
\begin{center}
\includegraphics[width=0.60\linewidth]{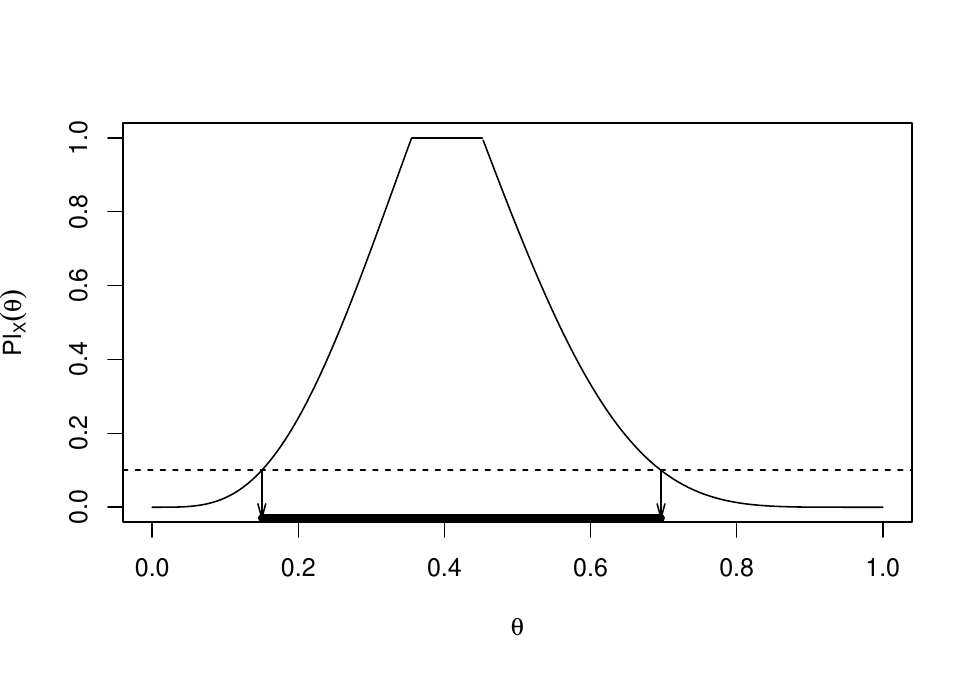}
\vspace{-0.15in}
\caption{The plausibility curve of the binomial example in Section \ref{ss:ims} for the case with $n=10$ and $x=4$. The 90\% plausibility interval is given by the points with $\mbox{Pl}_X(\theta)\geq 0.1$.
}
\label{fig:pl-curve}
\end{center}
\end{figure}

As a computational procedure, IMs consists of three steps, association, prediction, and combination. The association step represents the sampling model by \eqref{eq:binomial-ims}, which
associates the observed data $X$, the unknown parameter $\theta$, and the auxiliary variable $U$ that is missing but predictable. The prediction step predicts the missing value $U$ using a so-called valid predictive random set $\mathcal{S}$, resulting in \textit{epistemic probabilities}. For example, for a general purpose inference, we can take 
\[
\mathcal{S} = [U/2, 1-U/2],\qquad U\sim \mbox{Uniform}(0,1),
\]
which is valid \citep{martin2013inferential}. 
The combination step combines  the predictive random set $\mathcal{S}$ and the observed data to push $\mathcal{S}$ into a random set on $\Theta$ based on the association \eqref{eq:binomial-ims}:
\[
\Theta_{X}(\mathcal{S}) = \cup_{u\in \mathcal{S}}\Theta_X(u),
\]
where $\Theta_X(u)$ is defined in \eqref{eq:DS-set-value}.
For any assertion or hypothesis of interest $A$, this IM model produces the lower probability called {\it belief}
\[
\mbox{Bel}_X(A) = \mbox{Prob}(\Theta_{X}(\mathcal{S}) \subseteq A),
\]
the probability of the truth of $A$,
and the upper probability called {\it plausibility}
\[
\mbox{Pl}_X(A) = 1-\mbox{Prob}(\Theta_{X}(\mathcal{S}) \subseteq A^c),
\]
the plausibility for the truth of $A$. For example, for all $\theta\in [0,1]$, we have
\begin{eqnarray}
\mbox{Pl}_X(\{\theta\}) &=& 1-\mbox{Prob}(\Theta_{X}(\mathcal{S}) \subseteq \{\xi: 0\leq \xi\leq 1, \xi\neq \theta\}) \nonumber\\
&=& \left\{\begin{array}{lcl}
2\mbox{Beta}(\theta, X, n-X+1)),&&\mbox{if $\theta < \mbox{Beta}^{-1}(0.5, X, n-X+1)$;}\\
2(1-\mbox{Beta}(\theta, X+1, n-X)),&&\mbox{if $\theta > \mbox{Beta}^{-1}(0.5, X+1, n-X)$;}\\
1, && \mbox{otherwise.}
\end{array}
\right.
\label{eq:pl-binom}
\end{eqnarray}

\newcommand{\minp}[1]{\begin{minipage}{5.5in}#1\end{minipage}}
\newcommand{\mina}[1]{\begin{minipage}{2.85in}#1\end{minipage}}
\newcommand{\minc}[1]{\begin{minipage}{2.85in}#1\end{minipage}}

\begin{table}
\spacingset{1.0}\small
\begin{center}

\begin{tabular}{lcl}
\hline\hline
\multicolumn{3}{c}{}\\
\multicolumn{3}{c}{\textit{\textbf{ Properties Supporting IMs as a Generalized Logic of Science}}}\\
\cline{1-3}
\multicolumn{3}{c}{}\\
\multicolumn{3}{l}{\minp{ 
    $\checkmark$ Objective and Probabilistic Reasoning\\
    $\checkmark$ Handling Hypotheses and Uncertainty\\
    $\checkmark$ Inductive Reasoning\\
    $\checkmark$ Flexibility Across Domains\\
    $\checkmark$ Empirical Basis for Hypothesis Testing\\
    $\checkmark$ Bridging Bayesian and Frequentist Paradigms. }}\\
    \multicolumn{3}{c}{}\\
  \hline
   \multicolumn{3}{c}{}\\
\multicolumn{1}{c}{\textit{\textbf{Challenges and Open Questions}}} && \multicolumn{1}{c}{ \textit{\textbf{Future Directions}}}\\
    \cline{1-1}\cline{3-3}
\multicolumn{3}{c}{}\\

    \mina{\textcircled{1} Scalability\\
       \textcircled{2} Acceptance in the Scientific Community\\
       \textcircled{3} Extensions and Generalizations\\
       \textcircled{4} Philosophical Foundations}	&&
    \minc{
       \textcircled{4} Empirical Applications\\
       \textcircled{3} Theoretical Development\\
       \textcircled{2} Educational Efforts\\
       \textcircled{1} Integration with Computational Tools
       }\\   
\multicolumn{3}{c}{}\\
\hline
\end{tabular}    
\end{center}
\caption{The LLM-based computational TB-evaluation of IMs as a generalized logic of science.  }
\label{tbl:ChatGPT-ims}
\end{table}

IMs are probabilistic and have desirable frequency properties. In the context of significance testing, the belief and plausibility functions generate (situation-specific) probabilistic uncertainty assessments of hypotheses of interest. The IM counterpart of Neyman-Pearson confidence interval is plausibility interval, the collection of singleton assertions with plausibility \eqref{eq:pl-binom} greater than or equal to the given frequentist error rate $\alpha$. This is illustrated in Figure \ref{fig:pl-curve}, where the plausibility curve can be viewed as a upside-down confidence curve. It should be noted that, philosophically, unlike Neyman-Pearson's confidence intervals, plausibility intervals are appealing because while guarantee exact frequentist coverage, they offer situation-specific probabilistically interpretable uncertainty assessments. That is, plausibility intervals can be interpreted both freqentistly and Bayesianly.

Since the IMs framework is relatively new, an extensive TB-evaluation depends on future research. Here, we conduct an experiment of computational TB-evaluation using ChatGPT, with a manually dynamic chain of thought and verification approach \citep{wei2022chain,dhuliawala2023chain}. The detailed process and intermediate results are provided in \Supplementary~C. The key concluding results are summarized in Table \ref{tbl:ChatGPT-ims}, which we found reasonably meaningful and even valuable for future research, considering that the assessments are done at the language level. 
Since ChatGPT is generative, we will not over-interpret its results. Nevertheless, formal statistical and computational TB-evaluation is subject to future development.






\ifthenelse{1=1}{}{
\subsection{Resampling: 
from Jackknife to 
(Calibrated) Bootstrap
}\label{ss:bootstrap}

Reflecting on the advances in developing inferential methods for the logic of scientific reasoning, as we did in the last section under the term 'inferology,' we find that by the 1950s, Neyman-Pearson's frequentist framework—comprising confidence intervals for inference and the reject-or-fail-to-reject approach for hypothesis testing—was widely accepted in scientific practice. The critical techniques of maximum likelihood estimation (MLE) and the associated large-sample theory-based approximation were developed in a manner that might be described, as \cite{kuhn1970structure} calls it, as 'normal science.' In this context, the less-than-ideal finite-sample performance of MLE was recognized within the research community. This led to  weakened beliefs in Fisher's theory of MLE, and thus the problem to be solved was clearly established. From a scientific creativity perspective, as far as the authors can trace, this issue has since undergone several exploratory and transformational shifts. We outline below the key reactivities within the proposed weak belief framework for analyzing scientific reactivities, spanning developments from the {\it Jackknife} to the {\it Bootstrap} and to the most recent inquiries into {\it finite-sample exact inference}.

\begin{table}\color{blue}\spacingset{0.82}\small
\begin{tabular}{cclclcl}
\hline\hline
    Year &&	Event / Contribution	&& Authors	&& Significance / Notes\\
    \cline{1-1}\cline{3-3}\cline{5-5}\cline{7-7}
    \multicolumn{7}{c}{}\\
1969	&&\begin{minipage}{1.5in}
    Introduction of the Jackknife
\end{minipage} 	&&\mina{Tukey}&&	\minc{Laid foundational ideas for resampling techniques by systematically leaving out one observation at a time to estimate variance. Limited to linear statistics.}\\
\multicolumn{7}{c}{}\\
1979&&	\minp{Bootstrap Method Introduced}	&&\mina{Efron}	&&\minc{Efron introduced the bootstrap as a novel resampling approach to approximate the distribution of complex statistics, generalizing the jackknife to nonlinear and non-smooth cases.}\\
    \multicolumn{7}{c}{}\\
1981&&	\minp{Theoretical Asymptotics for Bootstrap}&&\mina{	Bickel and Freedman; Singh}&&\minc{	Demonstrated the theoretical foundations of bootstrap asymptotic consistency, which spurred its acceptance and practical applications.}\\
    \multicolumn{7}{c}{}\\
1984&&	\minp{Bayesian Bootstrap}	&&\mina{Rubin}	&&\minc{Expanded the bootstrap method to include Bayesian principles, enhancing its application scope by incorporating prior beliefs.}\\
    \multicolumn{7}{c}{}\\
1986	&&\minp{Approximate Bootstrap Confidence Intervals}&&	\mina{Efron and Tibshirani}	&&\minc{Developed methods to compute bootstrap-based confidence intervals, further advancing the practical use of bootstrap in statistical inference.}\\
    \multicolumn{7}{c}{}\\
1992&&	\minp{Bootstrap Applications to Regression and Time Series}	&&\mina{Davison and Hinkley} &&	\minc{Extended bootstrap methodology to complex models, popularizing its use in linear regression, time series, and related applications.}\\
    \multicolumn{7}{c}{}\\
1996&&	\minp{Bootstrap in High Dimensions}&&\mina{	Wellner and Zhan}&&	\minc{Investigated the limitations of bootstrap in high-dimensional cases, revealing challenges in situations with a large parameter-to-sample size ratio.}\\
    \multicolumn{7}{c}{}\\
2005&&	\minp{Bayesian Resampling}	&&\mina{Newton and Raftery}&&	\minc{Developed Bayesian resampling strategies for hierarchical models, offering a hybrid method combining bootstrap and Bayesian techniques.}\\
    \multicolumn{7}{c}{}\\
2018&&	\minp{Bootstrap Failures in High Dimensions}	&&\mina{El Karoui and Purdom}	&&\minc{Provided evidence that standard bootstrap can fail in high-dimensional settings, inspiring further developments in robust finite-sample methods.}\\
    \multicolumn{7}{c}{}\\
2020&&	\minp{Finite-Sample Valid Bootstrap Methods}	&&\mina{Various}&&	\minc{Emergence of methods addressing bootstrap’s limitations in finite samples, emphasizing exact or conservative inferences in high-dimensional and complex data contexts.}\\
    \multicolumn{7}{c}{}\\
------&&	\minp{Efficiency}	&&\mina{Various}&&	\minc{both computational (Wang and Martin, private conversations) and inferential (Conditional IMs).}    \\
    \multicolumn{7}{c}{}\\
\hline
\end{tabular}
\caption{[FIX THE TABLE CONTENT] Creativity observations, consisting of historical and creative events, in the development of the bootstrap resampling method}
\label{tbl:bootstrap}
\end{table}

\addText{
Table \ref{tbl:bootstrap}, a data table for historical and creative events in the development of the bootstrap resampling method, highlights key advances, authors, and the significance of each contribution
This table format captures both historical and innovative aspects, providing context and highlighting the bootstrap's evolution and adaptations to emerging challenges, especially in high-dimensional and finite-sample inference settings.
}

\subsubsection{Jackknife}

Jackknife marks the transformational change of considering what has been nowadays understood as cross-validation analysis by \cite{quenouille1949approximate,quenouille1956notes}
and the exploratory advance to considering variance estimations as well about a decade later by \cite{tukey1958bias}; See \cite{shao2012jackknife}, \cite{lahiri2013resampling}, and references therein for more discussion.
Suppose we have a sample $x=(x_1, ..., x_n)$ and an estimator $\hat{\theta} = s(x)$. The jackknife
focuses on the samples that {\it leave out one observation at a time}:
\[x_{(i)}=(x_1,...,x_{(i-1)},x_{(i+1)},...,x_n)\]
for $i=1,..., n$, called jackknife samples. The $i$-th jackknife sample consists of the data set
with the i$i$-h observation removed. Let $\hat{\theta}_{(i)}=s(x_{(i)})$ be the $i$-th jackknife replication of $\hat{\theta}$.
The jackknife estimate of sqaured standard error is defined as the sample variance of the jackknife replicates adjusted by the factor of $(n-1)$.

The jackknife resampling method proved effective primarily for linear statistics, such as the mean, but encountered limitations when applied to non-smooth statistics (e.g., the median) or nonlinear statistics (e.g., the correlation coefficient). The jackknife's accuracy issues with these cases stemmed from inconsistencies between predicted and observed values when applied to estimands beyond the mean. These limitations prompted the development of improved techniques, aimed at extending resampling accuracy across a broader range of statistical measures.

\subsubsection{Bootstrap}

To improve the performance of jackknife in the realm of statistical inference and prediction,
bootstrap methods are discovered to estimate the sample distribution of the statistic of interest by resampling the observed data. Since the seminal paper of \cite{efron1979bootstrap}, 
these methods have made the paradigm shit in the field of resampling methods \citep{efron1993introduction,shao2012jackknife,lahiri2013resampling}.
In addition to the development of bootstrap itself, it has soon motivated relevant research in other contexts, including its Bayesian counterpart \citep[see][and references therein]{rubin1981bayesian,efron1982jackknife,newton1994approximate,efron2012bayesian,newton2021weighted}.

Approximation methods often lack theoretical support for finite-sample problems, and this popular bootstrap is no exception. Theoretically, bootstrap methods primarily focus on asymptotic validity \citep[see, {\it e.g.},][ and references therein]{bickel1981some,singh1981asymptotic, wellner1996bootstrapping,vanderVaart1996weak,kosorok2008introduction,efron1993introduction}. However, even within large-sample contexts, the bootstrap can fail \citep[see, {\it e.g.,}][and references therein]{bickel1983bootstrapping,abrevaya2005bootstrap, chernozhukov2023high}. For instance, in high-dimensional linear regression where $n \to \infty$ and $p/n \to c > 0$ for $p$ predictors,  
the bootstrap has been found to yield inconsistent estimates of the true sampling distribution of regression coefficients, with poor numerical inference results for the vector of coefficients $\beta$  \citep{el2018can}. Most importantly, these works offer warnings about the application of bootstrap in high-dimensional settings. From the perspective of our \weak~belief framework, these {\it thought experiments} provide evidence as a way of verification, leading to further investigations.

To address limitations of the standard $n$-out-of-$n$ bootstrap method in certain cases, researchers have explored alternative resampling methods, such as $m$-out-of-$n$ resampling with or without replacement (known as subset sampling), originally suggested by \cite{bretagnolle1983lois} and further developed in works like \cite{politis1994large, bickel2008choice, bickel2012resampling}. Despite these efforts and their asymptotic backing, a significant gap remains in developing theoretically supported 
bootstrap methods for high-dimensional inference problems.

\subsubsection{Calibrated bootstrap}

Most recently, revolving the problem of bootstrap inavoidably demands finite-sample valid inference methods using resampling.
To achieve reliable finite-sample performance, \cite{jiang2024finite} proposes an efficient resampling-based strategy to closely approximate the theoretical solution. Their innovation lies in recognizing that resampling variability and inferential uncertainty are distinct concepts and that they consider an approach aiming to align resampling variability with inferential uncertainty. Using an inferential models (IMs) perspective, rooted in the pivotal method, they numerically assess uncertainty via loss functions in parameter estimation, applying cumulative distributions to introduce pivotal quantities. This leads to the method of calibrated bootstrap, a resampling method that matches the theoretical distribution for valid inference, making it suitable for high-dimensional cases where standard bootstrap methods are known to falter.

\footnote{See \cite{cella2024variational} etc?}
With the rapid advancements in statistical inferential methods driven by the needs of artificial intelligence, we may be on the cusp of a significant paradigm shift: the emergence of finite-sample exact inference. Unlike traditional inference, which often relies on asymptotic approximations that may fail in high-dimensional or complex models, finite-sample exact inference aims to provide rigorous, reliable results even with limited data. Such a paradigm could reshape not only AI and machine learning but also fields such as genomics, finance, and healthcare, where accurate, model-consistent inference from small or complex data sets is essential. In this context, developing computationally feasible methods with strong theoretical foundations becomes increasingly critical, laying the groundwork for a future where inference is both precise and robust, irrespective of sample size.
} 

\section{Concluding Remarks}\label{s:conclusion}

Currently, statistics is comprised of a multitude of disparate estimation and hypothesis testing techniques, and when a hypothesis is rejected, it's often unclear how to incorporate the knowledge of that rejection into the new estimation problem. However, in science, the formulation of hypotheses and their testing does not occur in a vacuum, and more plausible alternatives are often required to gain broad traction when rejecting an existing theory. In this work, we introduce an overarching framework that ties together hypothesis testing and estimation for the purposes of making scientific discoveries. This framework provides a direction to reconcile differences in foundational approaches to statistics, such as frequentism, Bayesianism, and inferential models, by placing them in their proper context within the process of creativity. While creation's natural statistical analogue is modeling and estimation and evaluation's is hypothesis testing, there are not yet well established methods for \articulation. In our current era of weak-AI development, we have the ability to estimate arbitrary functions and incorporate data of various modalities into prediction machines, but these are all constrained in their capacity by their formulation. Strong AI will need to encompass all three steps of scientific discovery, and improvements in the creation and \articulation~steps will be necessary to get there.

Technically, we proposed a very simple {\it statistical} framework for understanding and analyzing creativity \citep[see, {\it e.g.},][for more elaborated systems on general creativity]{wiggins2006preliminary}. This framework is inspired by a detailed study of significant discoveries in celestial mechanics and physics and is illustrated through both a straightforward artificial example and the extensive 260-year-long pursuit of a sound logic of science. Although we only conducted a limited computational experiment on TB-evaluation of IMs using an LLM with manually dynamic chain of thought and verification, computational TB-creation and TB-\articulation~can be considered in future research by making use of generative AI models and conducting automodeling, which has motivated relevant research \citep[see, {\it e.g.},][and references therein]{sun2024survey}.

However, we acknowledge that TB itself is built primarily on inductive reasoning, which inherently leaves it open to further evaluation and refinement. To enable such evaluations, it would be valuable to collect systematic data on scientific creativity. Such data would not only facilitate rigorous assessment of the proposed framework but also support its expansion, thereby establishing empirical foundations for computational methods to understand and analyze scientific creativity. In the long run, this could lead to methods capable of generating scientific ideas and solving complex problems. At this juncture, it is important to note that exciting progress has already been made in fields such as computational creativity and experimental creativity, where data-driven approaches and experimental observations provide crucial insights \citep[see, {\it e.g.}][]{varshney2019big,soroush2024loosely}. Leveraging these advancements can further strengthen the empirical grounding of our framework and enhance our understanding of the mechanisms behind scientific creativity.




\ifthenelse{1=1}{}{
{\color{magenta}
What said in \cite{sep-scientific-discovery} seems to apply here but with a modification for our contribution to the subtle topic of scientific creativity that can be potentially useful for development of strong AI:
 
\begin{quote}
  This essay has illustrated that philosophy of discovery has come full circle. Philosophy of discovery has once again become a thriving field of philosophical study, now intersecting with, and drawing on philosophical and empirical studies of creative thinking, problem solving under uncertainty, collective knowledge production, and machine learning. Recent approaches to discovery are typically explicitly interdisciplinary and integrative, cutting across previous distinctions among hypothesis generation and theory building, data collection, assessment, and selection; as well as descriptive-analytic, historical, and normative perspectives (Danks \& Ippoliti 2018, Michel 2021). The goal no longer is to provide one overarching account of scientific discovery but to produce multifaceted analyses of past and present activities of knowledge generation in all their complexity and heterogeneity that are illuminating to the non-scientist and the scientific researcher alike.  
\end{quote}

 \noindent
With this perspective, our three examples selected from the field of statistical science demonstrate systematic investigation at the level of studying innovative ideas, termed creatology and inferology, demands large scale collaboration among all scientific fields.\footnote{{\color{red}Check where to find ``data sources of creativity and innovation''.}}

Although the discovery of Neptune has been chosen as an illustrative example here in this paper, the are numerous significant creative and innovative discoveries in science since the ancient times that can be fit into the proposed framework. In the context of the Neptune example, for example,
the history of physics and astronomy alone is marked by numerous significant creative and innovative discoveries, from ancient times to the modern era.
To improve and enrich the proposed theory of weak beliefs for creativity and discovery, it will be helpful to examine such scientific discoveries, which have not only expanded humanity's understanding of the universe but also led to technological advancements that have transformed society. It is our expectation that such endeavors would lead to a sound theory of scientific creativity for the development of strong artificial intelligence powered with a certain degree of human wisdom.
}

\begin{comment}[TODO]
    \color{red}A summary of computational implementation? See Section 10 of \cite{sep-scientific-discovery}.
\end{comment}

\begin{comment}[TODO]
    \color{red} How can the proposed theory of \weak~beliefs be generalized\footnote{Is cognitivity include creativity or vice versa?} or how can a new framework be developed to cover general cognitivity (or cognitive abilities)?  Or, it can be anything meaningful can be said about string AI.
\end{comment}

}


\if0\blind
\section*{Acknowledgment}

{The authors are supported in part by U.S. National Science Foundation grant DMS-2412629.}
\fi

\if1\blind
\vspace{1.0in}
\fi

\bibliographystyle{apalike}
\bibliography{references}

\ifbool{is_arXiv}{
\pagebreak

\ifthenelse{1=1}{}{
\setcounter{page}{1}

\centerline{\huge Towards Strong AI: 
\Weak~Beliefs}

\vspace{0.1in}

\centerline{\huge  and Scientific Creativity}

\vspace{0.2in}

\centerline{\textbf{\textit{\Large --- Supplementary}}}
}

\centerline{\textbf{\textit{\Large Appendices: Supplementary Materials}}}

\appendix

\section{More examples of great innovations in astronomy and physics}

Far more historical examples are available than we have had space to exploit here.
In this section, we briefly summarize the observations of our studies of great discoveries in celestial mechanics and physics from a statistical perspective of their innovations. The first is the invention of the {\it geocentric model} that hypothesizes that the Sun, Moon, stars, and planets all orbit Earth. 
It was the predominant description of the cosmos in many ancient European civilizations, such as those of Aristotle (384–322 BC) in classical Greece and Ptolemy in Roman Egypt, as well as during the Islamic Golden Age. Claudius Ptolemy (c. 100 - c. 170 AD), an Alexandrian mathematician, astronomer, astrologer, geographer, and music theorist, thought the solar system looked like this and standardized geocentrism. From a statistical perspective of the underlying creativity, we note that this model was built on two observations \citep[][pp. 5-20  and Wikipedia, 10/23/2024]{kuhn1992copernican}: 
\begin{quote}\it First, from anywhere on Earth, the Sun appears to revolve around Earth once per day. While the Moon and the planets have their own motions, they also appear to revolve around Earth about once a day. The stars appeared to be fixed on a celestial sphere rotating once each day about an axis through the geographic poles of Earth. Second, the Earth seems to be unmoving from the perspective of an earthbound observer; it feels solid, stable, and stationary.
\end{quote}

The geocentric model dominated for centuries until the discovery of the heliocentric system, known as the {\it Heliocentric Hypothesis}, of the Greek astronomer and mathematician Aristarchus of Samos (c.310 – c. 230 BC). The heliocentric model places all of the then-known planets in their correct order around the Sun. 
With the invention of the telescope in 1609, {\it observations} made by Galileo Galilei led him to the thought that while this observation was incompatible with the Ptolemaic system, it was a natural consequence of the heliocentric system. Continued {\it observations} and analysis led to variations of
heliocentric system, where planets orbit in perfect circles.

Johannes Kepler (1571 – 1630) 
analyzed Tycho Brahe's 
famously accurate {\it observations} and afterwards constructed his three laws in 1609 and 1619: 
the orbit of a planet is an ellipse with the Sun at one of the two foci;
a line segment joining a planet and the Sun sweeps out equal areas during equal intervals of time; and
the square of a planet's orbital period is proportional to the cube of the length of the semi-major axis of its orbit.     
The application of these laws dramatically improved the accuracy of celestial {\it observations} and {\it predictions}. Violations of Kepler's laws, indicated by inconsistency between the observed and predicted for Mercury, 
have been used to explore more sophisticated models of {\it gravity}, such as general relativity. While Newton's laws generalize Kepler's laws, most problems related to the periods of orbits are still best solved using Kepler's laws, since they are simpler. 

Great discoveries were also found in classical physics. For example, the 17th-19th centuries saw Newton's laws of motion and universal gravitation (1687), 
electromagnetism, and thermodynamics.
In April 1820, Hans Christian Ørsted observed that an electrical current in a wire caused a nearby compass needle to move. 
This discovery represented a major step toward a unified concept of energy. This unification, which was observed by Michael Faraday, extended by James Clerk Maxwell, and partially reformulated by Oliver Heaviside and Heinrich Hertz, is one of the key accomplishments of 19th-century mathematical physics \citep{darrigol2003electrodynamics}. 
Scientists like Sadi Carnot, Lord Kelvin, and Rudolf Clausius laid the foundation for the study of heat, energy, and entropy, which had profound implications for understanding physical processes. Discoveries can often be well understood by tracing their history. 
Tracing their history, one would see that like the Neptune discovery, geocentric system, heliocentric system, Kepler's laws, Newton's laws, and Einstein's theoy of special and general relativity, such great discoveries all start with experiments and observations, build mathematical theories or statistical models, verify the new theories with experiments and new observations, and iterate such a process to further verification, improvement, and discoveries. 

Extensive existing research on the science of creativity and discovery, primarily conducted by philosophers of science, has also been undertaken \citep[see, {\it e.g.},][and references therein]{kuhn1970structure,Aleinikov2013,sep-scientific-discovery}. Their perceived general structure or pattern of such activities, particularly in natural science, exhibits similarities to our observations in this section.

\section{The implementation of the Gibbs sampler and the EM algorithm for the mixture of normals}\label{a:mnms-gibbs}

With $\mu_0 = 0$ and $\sigma_0^2 = 10^4$, the steps for a Gibbs sampler with a Dirichlet prior are
\begin{enumerate}
    \item Initialize $\pi^{(1)}_1, \dots, \pi_1^{(K)}$ and $\phi^{(1)}, \dots, \phi^{(K)}$ such that $\sum_{i=1}^K \pi_1^{(i)} = 1$ and $\phi_{i}^{(1)} \leq \phi_{i+1}^{(i)}$ for all $i \in 1, 2,\dots, K-1$.
    \item Sample the augmented variable $$Z^{(t)}_i\sim P(Z_i = k | \pi^{(t)}_1, \dots, \pi^{(t)}_K, \phi^{(t)}_1, \dots, \pi^{(t)}_K) \propto \pi^{(t)}_{k} \exp\left(-\frac{(Y_i - \phi_i^{(t)})^2}{2}\right)$$
    for $i = 1, \dots, n$.
    \item Using $\eta_k^{(t)} = \sum_{i=1}^n \mathbb{I}(Z_i^{(t)} = k)$ for $k = 1,\dots, K$, sample 
    $$(\pi^{(t+1)}_1, \dots, \pi^{(t+1)}_K) \sim \text{Dirichlet}(1 + \eta_1^{t}, \dots, 1+\eta_K^{(t)})$$
    \item Let $\bar \phi_k = \frac{1}{\eta_k^{(t)}} \sum_{i=1}^n Y_i\mathbb{I}(Z_i^{(t)} = k)$, and sample the posterior means
    $$\phi_i^{(t+1)} \sim N\left(\frac{\frac{\mu_0}{\sigma^2_0}+ \eta_k^{(t)}\bar\phi_k^{(t)}}{\frac{1}{\sigma^2_0} + \eta_k^{(t)}}, \left(\frac{1}{\sigma^2_0} + \eta_k^{(t)}\right)^{-1}\right)$$
    and go to step 2 to repeat.
\end{enumerate}
If instead, we wish to calculate the MLE for the initial estimate, for brevity, we present that case when $K_n=2$, where the $k^{th}$ expectation step is

$$w_i^{(k)} = \frac{\hat\pi_1^{(k-1)}\exp\left(-(y_i-\hat\phi_1^{(k)})^2\right)}{\hat\pi_1^{(k-1)}\exp\left(-(y_i-\hat\phi_1^{(k)})^2\right)+\left(1-\hat\pi_1^{(k-1)}\right)\exp\left(-(y_i-\hat\phi_2^{(k-1)})^2\right)},\qquad i = 1,\dots,n$$
and the $k^{th}$ maximization step is
$$\hat\pi_1^{(k)} = \frac{1}{n}\sum_{i=1}^n w_i^{k},\quad \hat\phi_1^{(k)} = \frac{\sum_{i=1}^n w_i^{(k)}Y_i}{\sum_{i=1}^n w_i^{(k)}},\quad\text{and}\quad \hat\phi_2^{(k)} = \frac{\sum_{i=1}^n (1-w_i^{(k)})Y_i}{\sum_{i=1}^n (1-w_i^{(k)})}.$$
By choosing some reasonable initialization values, such as $\pi_1^{(0)} = 0.5$, $\phi_1^{(0)} = \bar{Y} - 1$, and $\phi_2^{(0)} = \bar{Y}-1$, and iterating until the $\|\hat{\pi}^{(k)} - \hat{\pi}^{(k-1)}\|<10^6$, we obtain an approximate value for the MLE of the model parameters. The analogous setup where $K_n>2$ has a separate $w_i$ calculated for each component to estimate a larger set of $\pi_i$ values. Then, the $\phi_i$ are calculated similarly to above for each of the components.


\section{A computational TB-evaluation with ChatGPT via a chain-of-thought and verification approach} 

We conducted a simple experiment of computational creativity to create {\it a generalized logic of science} with ChatGPT \citep{openai2024chatgpt}, a large language model. We use the chain-of-thought mechanism \citep{wei2022chain,dhuliawala2023chain} in place of TB-evaluation. 
While future of development will involve automated reverse-of-engineering methods, in this experiment we manually provide a chain of prompts in terms of the chain-of-thought. First, potential candidates are found to satisfy the criteria that it has to be probabilistic and that it has to produce frequency-calibrated probabilistic assessments of hypotheses and assertions. These are presented in Section \ref{ss:comp-create-desirabl-logic}, which gives the following candidates:
\begin{itemize}
    \item Fiducial Inference
    \item Inferential Models (IMs)
    \item Calibrated Bayesian Methods
    \item Confidence Posterior Distributions
\end{itemize}
While Fiducial Inference is not frequency-calibrated, one may follow Calibrated Bayesian Methods and Confidence Posterior Distributions.
However, in responding to ChatGPT's question ``Would you like a deeper dive into any of these, or should we explore potential applications?'' a deeper dive into into IMs is presented in Section \ref{ss:comp-create-IMs}. This deeper diver provides a type of TB evaluation. 
The evaluation summarizes the following properties to support IMs as a potential generalized logic of science:  
    objective and probabilistic reasoning,
    handling hypotheses and uncertainty,
    inductive reasoning,
    flexibility across domains,
    empirical basis for hypothesis testing, and
    bridging bayesian and frequentist paradigms.

Most important, perhaps, it shows future evaluations, expressed in terms of   
\begin{itemize}
    \item Challenges and Open Questions
    \begin{itemize}
       \item Scalability
       \item Acceptance in the Scientific Community
       \item Extensions and Generalizations
       \item Philosophical Foundations
    \end{itemize}
    \item Future Directions
    \begin{itemize}
       \item Empirical Applications
       \item Theoretical Development
       \item Educational Efforts
       \item Integration with Computational Tools
    \end{itemize}
\end{itemize}
with the final conclusion:
\begin{quote}\it
    Inferential Models (IMs) represent a promising candidate for a generalized logic
of science, combining probabilistic reasoning with rigorous frequency calibration.
By bridging Bayesian and frequentist paradigms and avoiding subjective priors,
IMs offer a powerful framework for objective and inductive scientific reasoning.
Their success in achieving widespread adoption will depend on overcoming computational
and theoretical challenges while demonstrating their value in diverse scientific
contexts.
\end{quote}


\subsection{Find inferential frameworks as potential candidates}\label{ss:comp-create-desirabl-logic}


\noindent
\subsubsection*{Prompt 1$>$ Consider the development of the logic of science. How much do you know about the history of efforts in this area?}

\noindent
\textbf{ChatGPT}:: The development of the logic of science has a rich history, shaped by contributions from philosophers, mathematicians, and scientists seeking to understand and formalize how scientific reasoning works. Here's a brief overview of key milestones:

\noindent
\textit{Early Foundations}
\begin{enumerate}
    \item[1.] Aristotle (384-322 BCE): Laid the groundwork with syllogistic logic, offering a system for deductive reasoning that influenced scientific thought for centuries.
    \item[2.] Islamic Golden Age (8th-13th centuries): Scholars like Alhazen emphasized empirical observation and experimentation, contributing to the foundations of the scientific method.
\end{enumerate}
\noindent
\textit{Modern Logic and Science}
\begin{enumerate}
    \item[3.] Francis Bacon (1561-1626): Advocated for inductive reasoning in his work Novum Organum, marking a departure from purely deductive methods.
    \item[4.] René Descartes (1596-1650): Promoted deductive reasoning and systematic doubt, integrating mathematics with scientific inquiry.
\end{enumerate}
\noindent
\textit{The Emergence of Formal Logic}
\begin{enumerate}
    \item[5.] Immanuel Kant (1724-1804): Explored the limits of human knowledge, differentiating between a priori and a posteriori reasoning.
    \item[6.] 19th Century Advances: George Boole and others developed symbolic logic, paving the way for a more formal analysis of scientific reasoning.
\end{enumerate}    
    
\noindent
\textit{20th Century and Beyond}
\begin{enumerate}
    \item[7.] Karl Popper (1902-1994): Introduced falsifiability as a criterion for scientific theories, emphasizing critical testing over verification.
    \item[8.] Thomas Kuhn (1922-1996): Proposed the concept of paradigms and scientific revolutions in The Structure of Scientific Revolutions, highlighting the non-linear nature of scientific progress.
    \item[9.] Contemporary Efforts: Researchers in computational logic and AI are now exploring algorithmic approaches to scientific discovery and creativity.
\end{enumerate}
    
\noindent
How much detail are you interested in? I’d be happy to dive deeper into any specific era or figure!

\subsubsection*{Prompt 2$>$ Is Bayesian inference relevant in the logic of science?}

\textbf{ChatGPT}:: Yes, Bayesian inference plays a significant role in the logic of science, particularly in the way we model scientific reasoning and update beliefs in light of new evidence. Here's how Bayesian inference is relevant:
\begin{enumerate}
\item \textit{Modeling Scientific Reasoning}

Bayesian inference provides a formal framework for reasoning under uncertainty, which is central to scientific discovery. Scientists often begin with hypotheses (prior beliefs) and update these beliefs based on experimental data (evidence). This process aligns closely with Bayesian principles:
\[P(\mbox{Hypothesis}|\mbox{Data}) = P(\mbox{Data}|\mbox{Hypothesis})\cdot P(\mbox{Hypothesis})P(\mbox{Data}).\]

In this context:
\begin{quote}
    Prior $P(\mbox{Hypothesis})$: Initial belief about the hypothesis before observing data.\\
    Likelihood P(Data|Hypothesis): The probability of the data given the hypothesis.\\
    Posterior P(Hypothesis|Data): Updated belief about the hypothesis after observing data.
\end{quote}
\item \textit{Connecting Induction and Deduction}

Bayesian inference bridges the gap between inductive and deductive reasoning:
\begin{itemize}
    \item[]
    Induction: Updating probabilities based on observed data.
    \item[] Deduction: Using Bayesian models to derive consequences of a hypothesis.
\end{itemize}
This dual nature makes Bayesian methods a powerful tool in the logic of science, where both inductive learning and deductive validation are crucial.

\item \textit{Hypothesis Testing and Model Selection}

Bayesian inference provides a probabilistic approach to comparing models or hypotheses. For instance, Bayes factors allow scientists to evaluate the relative likelihood of competing theories. This is especially useful in fields like astrophysics, neuroscience, and machine learning, where competing models are common.

\item  \textit{Scientific Paradigms and Bayesian Updates}

In the context of Kuhn’s scientific paradigms, Bayesian inference explains how scientists update their beliefs when anomalies arise. As data accumulates, the posterior probability of a paradigm may shift, potentially triggering a paradigm shift when an alternative explanation becomes more plausible.

\item  \textit{Bayesian Philosophy of Science}

Philosophers like Edwin Jaynes and Richard Jeffrey have used Bayesian reasoning to argue that the scientific method can be interpreted as a process of Bayesian updating. This perspective emphasizes the probabilistic nature of scientific knowledge and its inherent uncertainty.
\end{enumerate}

\subsubsection*{Limitations and Criticisms}
\begin{description}
    \item[] \textit{Subjective Priors:} Critics argue that the choice of prior probabilities can be subjective and may influence results.
    \item[] \textit{Computational Complexity:} Bayesian methods can be computationally intensive for complex models.
\end{description}

Despite these challenges, Bayesian inference remains a cornerstone of modern scientific methodology, offering a formalized approach to updating beliefs and making predictions in an uncertain world.

Would you like to explore specific examples of Bayesian inference in scientific fields?

\subsubsection*{Prompt 3$>$ Are there prior-free alternatives that can be considered as the logic of science?}

\textbf{ChatGPT}:: Yes, there are alternatives to Bayesian inference that do not rely on specifying a prior, and these approaches can also be considered part of the logic of science. These "prior-free" methods aim to address concerns about subjectivity or dependence on prior information in Bayesian reasoning. Here are some notable examples:
\begin{enumerate}
    \item \textit{Frequentist Inference}

Frequentist methods dominate traditional statistical approaches and avoid prior probabilities by focusing solely on data and sampling distributions.
\begin{itemize}
    \item[]
    Core Idea: Hypothesis testing and parameter estimation are based on long-run frequency properties of data.
    \item[]
    Techniques:
        Maximum Likelihood Estimation (MLE): Determines parameter values that maximize the likelihood of observed data.
        Confidence Intervals: Provide a range of plausible values for a parameter without requiring prior beliefs.
        Null Hypothesis Significance Testing (NHST): Tests whether observed data is consistent with a null hypothesis.
\end{itemize}
Frequentist methods rely heavily on the data at hand and sidestep the need for subjective priors. However, they often lack a formal mechanism for incorporating prior information when it is available.

\item \textit{Likelihood-Based Inference}

Likelihood-based methods use the likelihood function to evaluate hypotheses or models without involving priors.

    Core Idea: Compare models or hypotheses directly based on how well they explain the observed data.
    Techniques:
        Likelihood Ratio Tests: Compare the likelihoods of two competing hypotheses.
        Akaike Information Criterion (AIC): Selects models based on their goodness of fit while penalizing for complexity.

This approach is more flexible than frequentist methods and avoids the subjectivity of prior selection, but it lacks the posterior probability interpretation that Bayesian inference offers.

\item \textit{Information-Theoretic Approaches}

Information theory offers another framework for the logic of science, focusing on how data reduces uncertainty.

    Core Idea: Measure and compare the information content of competing models.
    Examples:
        Minimum Description Length (MDL): Selects models that compress data most efficiently, balancing simplicity and explanatory power.
        Entropy-Based Methods: Quantify uncertainty reduction without explicit priors.

Information-theoretic approaches emphasize objectivity and are particularly useful in machine learning and model selection.

\item \textit{ Inductive Logic and Logical Positivism}

Philosophical frameworks for science that avoid priors include:

    Deductive-Nomological Model: Explains phenomena using general laws and initial conditions, avoiding probabilistic priors altogether.
    Carnap’s Logical Probability: An early attempt to formalize induction without strong priors, although later criticized for still requiring some subjectivity.

\item \textit{Empirical Bayes}

Empirical Bayes methods use data to estimate the prior, effectively making the prior less subjective.

    Core Idea: The prior is derived from the data itself, creating a hybrid between frequentist and Bayesian methods.
    Limitations: While this removes prior subjectivity, it introduces the risk of overfitting the prior to the observed data.

\item \textit{Pure Data-Driven Approaches}

Some emerging frameworks attempt to bypass priors entirely by relying solely on large datasets and computational methods:

    Machine Learning and AI: Algorithms like deep learning operate without explicit priors, relying on patterns in large-scale data.
    Symbolic Reasoning Systems: Use rule-based approaches to infer conclusions directly from observed patterns and logical relationships.

\item \textit{Falsification and Critical Rationalism}

Karl Popper’s philosophy emphasizes falsifiability as the hallmark of scientific reasoning.

    Core Idea: Science progresses by proposing bold hypotheses and subjecting them to rigorous testing. Hypotheses are retained only if they withstand attempts to falsify them.
    Strengths: Avoids reliance on probabilistic priors by focusing on hypothesis testing.
    Limitations: Does not provide a mechanism for assigning degrees of belief to hypotheses.
\end{enumerate}

\noindent
\textbf{Summary}

Free-prior alternatives, like frequentist inference, likelihood-based methods, information-theoretic approaches, and falsificationism, all offer viable frameworks for the logic of science. Each approach has strengths and limitations, and the choice of method often depends on the specific context of the problem being addressed.

Would you like to dive deeper into any of these alternatives?

ChatGPT can make mistakes. Check important info.

\subsubsection*{Prompt 4$>$ None of what you gave is probabilistic and produces frequency-calibrated uncertainty assessments. Do you know any that is probabilistic and produces frequency-calibrated uncertainty assessments?}

\textbf{ChatGPT}:: Ah, I see now. You're asking specifically for probabilistic methods that avoid subjective priors but still produce frequency-calibrated uncertainty assessments. This is a more nuanced question, but there are approaches that meet these criteria. Below are some probabilistic frameworks that aim for this balance:

\begin{enumerate}
    \item \textit{ Jeffreys' Priors (Objective Bayesian Methods)}

Jeffreys' priors are designed to be non-informative and invariant under reparameterization, providing a more "objective" approach within the Bayesian framework.

    Core Idea: These priors are derived mathematically from the Fisher information of the likelihood function and aim to reflect ignorance without subjective input.
    Frequency Calibration: Jeffreys' priors often produce posterior distributions with frequency-calibrated properties in well-behaved models, particularly for large sample sizes.
    Limitations: They can still exhibit pathological behavior in complex models or high dimensions.

\item \textit{ Reference Priors (Bernardo and Berger)}

Reference priors are another class of objective Bayesian priors specifically designed to maximize the influence of the data while minimizing the influence of the prior.

    Core Idea: They are constructed to maximize the divergence between the posterior and the prior, ensuring the posterior reflects information primarily from the data.
    Frequency Calibration: These priors often yield posterior credible intervals that align with frequentist confidence intervals asymptotically.
    Limitations: The construction process can be complex and depends on the problem's dimensionality and parameterization.

\item \textit{ Posterior Predictive Distributions (Cross-Validation Priors)}

Posterior predictive approaches avoid fixed priors by incorporating a form of cross-validation or empirical information.

    Core Idea: The prior is informed by partitioning the data, using part of the data to estimate the prior and the rest to compute the posterior.
    Frequency Calibration: By using data-driven priors, this method can provide frequency-aligned uncertainty estimates, particularly in repeated-sampling scenarios.
    Limitations: The results depend on how the data is split and may introduce subtle biases.

\item \textit{ Bootstrap Bayesian Inference}

The Bayesian bootstrap creates a posterior distribution without requiring a prior by resampling data.

    Core Idea: Instead of specifying a prior, resampling weights are treated probabilistically, generating a posterior distribution for the parameter of interest.
    Frequency Calibration: The resampling ensures that posterior intervals are frequency-calibrated in many practical scenarios.
    Limitations: It assumes the observed data distribution approximates the true distribution and can be sensitive to outliers or small sample sizes.

\item \textit{ Default or Improper Priors}

Improper priors (e.g., uniform priors over an infinite range) are another common choice for avoiding subjective inputs. While not technically "probabilistic" in the strict sense, they often yield posteriors that are well-calibrated in a frequentist sense.

    Core Idea: Use improper priors to focus entirely on the likelihood-driven posterior.
    Frequency Calibration: These approaches often yield posterior intervals with frequentist coverage properties.
    Limitations: Improper priors can sometimes lead to improper or undefined posteriors in complex models.

\item \textit{ Likelihood Weighting Methods}

Probabilistic methods based entirely on the likelihood function can produce uncertainty estimates without priors.

    Core Idea: Generate posterior-like distributions by normalizing likelihoods without explicitly defining a prior.
    Frequency Calibration: These distributions often retain calibration properties, particularly for large samples.
    Limitations: The approach lacks a rigorous probabilistic interpretation in the absence of a prior.

\item \textit{ SafeBayes (Vovk et al.)}

SafeBayes is a framework that adjusts Bayesian updating rules to ensure robustness and frequentist calibration, particularly in the presence of model misspecification.

    Core Idea: Modify the Bayesian updating process to prevent overconfidence, relying on likelihood-driven adjustments rather than subjective priors.
    Frequency Calibration: SafeBayes ensures that uncertainty estimates align with frequentist coverage properties, even under model misspecification.
    Limitations: The method is computationally intensive and less commonly used.
\end{enumerate}

\noindent\textbf{Summary}

If you're looking for probabilistic methods that produce frequency-calibrated uncertainty assessments, Jeffreys' priors, reference priors, and Bayesian bootstrap are strong candidates. These approaches strive to maintain the probabilistic nature of Bayesian inference while addressing concerns about subjectivity and ensuring frequentist calibration.

Would you like to explore any of these methods in more depth?

\subsubsection*{Prompt 5$>$ No, no, none of them satisfies the criteria. Please note, it has to be probabilistic, and the probabilistic statements on hypotheses or assertions have to be frequency-calibrated.  Would you like to give it another try?}

\textbf{ChatGPT}::  Thank you for clarifying the criteria. To meet the requirements, we need a framework that is probabilistic (assigns probabilities to hypotheses or assertions) and ensures that these probabilistic statements are frequency-calibrated (the stated probabilities correspond to long-run frequencies in repeated experiments). Here’s a revised attempt with a focus on approaches that truly satisfy both conditions:

\begin{enumerate}
    \item \textit{Hypothesis Testing with Randomized Confidence Distributions}

A confidence distribution is a frequentist probabilistic construct that assigns a distribution to a parameter in a way analogous to a Bayesian posterior, but without a prior. When extended to randomized confidence distributions, probabilistic assertions about hypotheses can achieve frequency calibration.

    Core Idea: Use randomized procedures to create a distribution over hypotheses or assertions such that, in the long run, stated probabilities align with observed frequencies.
    Frequency Calibration: These distributions are explicitly designed to provide probabilistic assessments that are calibrated in the frequentist sense.
    Practical Use: While not widely used, methods like fiducial inference (see below) can approximate this framework.

\item \textit{ Fiducial Inference (Fisher’s Fiducial Argument)}

Fiducial inference, originally proposed by R.A. Fisher, assigns a probabilistic distribution to a parameter based on the observed data, without requiring a prior.

    Core Idea: Transform data via the likelihood and pivot methods to derive probabilistic statements about parameters or hypotheses.
    Frequency Calibration: Fiducial distributions are designed to be frequency-calibrated by construction, ensuring that the probabilistic assertions match observed frequencies in repeated experiments.
    Limitations: Fiducial inference is not universally applicable and can break down in complex models.

\item \textit{ Calibration-Based Bayesian Methods (Calibrated Bayes)}

Calibrated Bayes methods modify Bayesian posterior distributions to ensure frequentist calibration of the resulting probabilistic statements.

    Core Idea: Adjust priors, likelihoods, or posterior probabilities using calibration techniques to align with long-run frequency properties.
    Frequency Calibration: These methods guarantee that posterior credible intervals and probabilities correspond to frequentist coverage probabilities.
    Examples:
        Posterior Predictive Checks: Incorporate posterior predictive simulations to ensure the probabilities reflect actual data-generating properties.
        Calibrated Credible Intervals: Explicitly adjust credible intervals to meet frequentist coverage requirements.

\item \textit{ Inferential Models (IMs)}

Proposed by Martin and Liu, Inferential Models are a modern extension of fiducial inference. They avoid priors but generate belief functions and plausibility functions that are probabilistically interpretable and frequency-calibrated.

    Core Idea: Use the observed data and a predictive random set to construct belief/plausibility functions, which serve as calibrated probabilistic measures.
    Frequency Calibration: By construction, belief probabilities satisfy frequency-calibration criteria in repeated sampling.
    Advantages: IMs can handle a wide variety of models while maintaining rigorous calibration properties.

\item \textit{ Confidence Posterior Distributions}

These are distributions over parameters that mimic Bayesian posteriors but are constructed entirely from data (no priors) and maintain frequentist properties.

    Core Idea: Derive distributions based on sampling distributions and pivot quantities that yield posterior-like intervals and probabilities.
    Frequency Calibration: The probabilities assigned to assertions about parameters or hypotheses are calibrated to match long-run frequencies.

\item \textit{ Default Bayes with Objective Calibration (Frequentist-Bayesian Reconciliation)}

This approach seeks to combine Bayesian and frequentist principles by choosing default priors or posterior adjustments to ensure frequency calibration.

    Core Idea: Modify the Bayesian framework to guarantee that posterior probabilities align with frequentist properties, such as coverage.
    Frequency Calibration: Guarantees that posterior intervals or probabilities meet desired frequentist criteria.
    Example: Using data-dependent priors designed to achieve coverage properties (e.g., Welch and Peers' matching priors).
\end{enumerate}

\noindent
\textbf{Why Most Approaches Fail}

Many probabilistic frameworks fail because:

    Bayesian methods prioritize coherence over calibration.
    Frequentist methods prioritize calibration but avoid assigning probabilities to hypotheses.
    Few frameworks integrate both probabilistic reasoning and calibration rigorously.

\noindent\textbf{Summary}

The most promising candidates that are probabilistic and yield frequency-calibrated uncertainty assessments include:
\begin{itemize}
    \item Fiducial Inference
    \item Inferential Models (IMs)
    \item Calibrated Bayesian Methods
    \item Confidence Posterior Distributions
\end{itemize}

Would you like a deeper dive into any of these, or should we explore potential applications?

\subsection{A deeper dive into IMs}\label{ss:comp-create-IMs}

\subsubsection*{Prompt 6$>$  Yes, I'd like a deeper dive into IMs, regarding its potential to become the generalized logic of science.}

\noindent
\textbf{ChatGPT}:: Inferential Models (IMs) as a Framework for the Logic of Science

Inferential Models (IMs), developed by Ryan Martin and Chuanhai Liu, offer a novel probabilistic framework that aligns closely with both the epistemic needs of scientific inquiry and the frequency-calibrated requirements of statistical reasoning. They hold potential as a generalized logic of science, particularly for probabilistic reasoning under uncertainty, with emphasis on inductive inference.

Here’s a deep dive into the IM framework and its potential role in the logic of science:

\subsubsection*{Key Features of IMs}
IMs operate by constructing belief functions and plausibility functions for making probabilistic assertions about hypotheses or parameters. These functions are rigorously calibrated in a frequentist sense, ensuring their probabilistic statements about assertions match observed frequencies in repeated experiments.

\begin{enumerate}
    \item \textit{Structure of IMs}

The IM framework consists of three core steps:
\begin{description}
    \item[\it Association:]
        Link the data, parameters (unknowns), and auxiliary random variables using a probabilistic model. This step formalizes the connection between observable and unobservable quantities, similar to a likelihood function in Bayesian inference but avoids priors.
        Example: $Y=g(\theta,U)$, where $Y$ is data, $\theta$ is the parameter, and $U$ is an auxiliary random variable.
    \item[\it Prediction:]
        Specify a random set for the auxiliary variable $U$, which encodes uncertainty in a way analogous to a prior but is derived entirely from the data-generating process. This step ensures objectivity.
    \item[\it Inference:]
        Use the association and the predictive random set to construct belief and plausibility functions, which provide calibrated probabilistic measures for hypotheses or parameter values.
\end{description}

\item \textit{ Belief and Plausibility}

    Belief Function $\mbox{Bel}(A)$: Quantifies the evidence in favor of a hypothesis $A$ based on the observed data.
    Plausibility Function $\mbox{Pl}(A)$: Quantifies the extent to which the data do not contradict $A$.
    The calibration of these functions ensures that their probabilistic statements about hypotheses have frequentist validity.

\item \textit{ Calibration}

IMs achieve frequency calibration by ensuring that the plausibility regions have proper coverage probabilities in repeated-sampling experiments. This is a fundamental property that many Bayesian methods lack due to their reliance on priors.
\end{enumerate}

\subsubsection*{Potential to Generalize the Logic of Science}

IMs naturally align with several foundational aspects of the logic of science:

\begin{enumerate}
    \item \textit{Objective and Probabilistic Reasoning}

    IMs avoid the need for subjective priors, offering a probabilistic framework derived solely from the data and model structure.
    Probabilistic statements are inherently objective and calibrated, addressing long-standing criticisms of subjectivity in Bayesian inference.

\item \textit{Handling Hypotheses and Uncertainty}

    Scientific reasoning often requires quantifying uncertainty about competing hypotheses. IMs provide a coherent mechanism for assigning belief and plausibility to scientific assertions, which can adapt to a wide range of contexts (e.g., causal inference, model selection).

\item \textit{Inductive Reasoning}

    Unlike purely deductive systems, IMs are inductive, focusing on learning from data and updating beliefs in light of evidence, a cornerstone of scientific inquiry.

\item \textit{Flexibility Across Domains}

    IMs can handle various kinds of models and data structures, including:
    \begin{itemize}
        \item[] Parametric and nonparametric settings.
        \item[] Complex models with auxiliary variables.
    \end{itemize}     
    This flexibility makes them suitable for generalizing across scientific domains.

\item \textit{Empirical Basis for Hypothesis Testing}

    IMs provide a robust foundation for hypothesis testing by:
    \begin{itemize}
        \item[] Defining plausibility regions for parameter values or hypotheses.
        \item[] Guaranteeing frequentist-calibrated error control.
    \end{itemize}
    This property aligns IMs with Karl Popper’s falsifiability principle, where hypotheses are rigorously tested against empirical evidence.

\item \textit{Bridging Bayesian and Frequentist Paradigms}

    IMs reconcile Bayesian-like probabilistic reasoning with frequentist calibration, potentially uniting the two paradigms in a generalized framework for scientific logic.
\end{enumerate}

\noindent
\subsubsection*{Challenges and Open Questions}

While IMs have great potential, several challenges must be addressed for them to fully mature as a generalized logic of science:
\begin{quote}
\begin{description}
    
    \item[\it Scalability:]
        Computational complexity can become an issue for high-dimensional models or large datasets.
    \item[\it Acceptance in the Scientific Community:]
        The novelty of the framework means it requires greater adoption and validation across diverse scientific fields.
    \item[\it Extensions and Generalizations:]
        IMs need further development to handle dynamic systems, time series, and other non-standard problems.
    \item[\it Philosophical Foundations:]
        The interpretation of belief and plausibility functions in the context of long-standing debates about probability (e.g., frequentist vs. subjective) needs more clarity.
\end{description}
\end{quote}

\subsubsection*{Future Directions}

To realize the potential of IMs as the logic of science, the following steps are crucial:
\begin{quote}
\begin{description}
    
    \item[\it 
    Empirical Applications:]
        Demonstrate the utility of IMs in solving real-world scientific problems, particularly in areas like physics, biology, and social sciences.
    \item[\it Theoretical Development:]
        Extend the framework to cover broader classes of models and inference problems, including causal inference and machine learning.
    \item[\it Educational Efforts:]
        Increase awareness and understanding of IMs through academic training and publications.
    \item[\it Integration with Computational Tools:]
        Develop software libraries to make IMs accessible for practical use.
\end{description}
\end{quote}

\subsubsection*{Conclusion}

Inferential Models (IMs) represent a promising candidate for a generalized logic of science, combining probabilistic reasoning with rigorous frequency calibration. By bridging Bayesian and frequentist paradigms and avoiding subjective priors, IMs offer a powerful framework for objective and inductive scientific reasoning. Their success in achieving widespread adoption will depend on overcoming computational and theoretical challenges while demonstrating their value in diverse scientific contexts.

Would you like to discuss specific applications of IMs or delve into the mechanics of their calibration and plausibility functions?


}{}

\end{document}